\documentclass[twocolumn,amsmath,amssymb,floatfix,nature,shownopacs,footinbib,superscriptaddress]{revtex4-2}
\usepackage{amsmath,amssymb,natbib,bm,epsf,color}
\usepackage{hyperref}
\usepackage{dblfloatfix}
\usepackage{etoolbox}
\appto\UrlBreaks{\do\-}
\usepackage{graphicx}
\usepackage{color}
\usepackage[utf8]{inputenc}
\usepackage[T1]{fontenc}
\usepackage{booktabs,array,longtable}
\usepackage{float}

\graphicspath{{images/}{supp_images/}}

\begin{document}

\title{Mobility energetic balance frames spatial configuration of cities}

\author{Federico Bellisardi}
\email{Corresponding authors: \\fbellisardi@ifisc.uib-csic.es}
\affiliation{Instituto de F\'{\i}sica Interdisciplinar y Sistemas Complejos IFISC (CSIC-UIB), Campus UIB, 07122 Palma de Mallorca, Spain}

\author{Enrique Cano Su\~n\'en}
\affiliation{Human Openware Research Lab (HOWLab), Engineering Research Institute of Aragon (I3A), University of Zaragoza (UZ), 50018 Zaragoza, Spain}

\author{Ana Ruiz-Varona}
\affiliation{Arquitecturas Open Source (AOS) Research Group, School of Architecture and Technology, Universidad San Jorge, 50830 Zaragoza, Spain}

\author{Jos\'e J. Ramasco}
\email{jramasco@ifisc.uib-csic.es}
\affiliation{Instituto de F\'{\i}sica Interdisciplinar y Sistemas Complejos IFISC (CSIC-UIB), Campus UIB, 07122 Palma de Mallorca, Spain}

\maketitle

{\bf Urban spatial configurations emerge as the result of interactions between citizens' activities, movements and the landscape. Topography forms part of the physical settings through which mobility occurs, influencing movement energetic requirements. Using road networks, elevation data and mobility flows inferred from geolocated records, we define a simplified comparative measure of movement energy and test how it changes when the observed urban configuration is translated, rotated or rescaled within its topographic setting while mobility demand is held fixed. Across 56 cities, observed configurations consistently show lower movement energy requirements than most of the alternative configurations examined. This recurrent pattern reveals a measurable relationship between topography and urban spatial organization and is compatible with cumulative adaptation among urban structure, mobility and terrain. These findings add an energetic dimension to our understanding of how topography relates to urban spatial organization and provide an empirical basis for considering terrain-related movement requirements when evaluating future urban growth and densification.}

Cities are complex systems whose spatial configuration emerges from the interaction of geographic, economic, social and infrastructural processes \cite{batty2013new,batty2008size,bettencourt2007growth,bassolas2019hierarchical,barthelemy2016structure}. Here, these configurations refer to how street networks, built areas, activities and mobility are organized within the physical and institutional conditions in which cities develop \cite{kansky1963structure,batty1994fractal,porta2006network,crucitti2006centrality,cardillo2006structural,strano2013urban}. Comparative analyses have identified distinct morphological types that capture the diversity of urban spatial configurations worldwide \cite{taubenbock2020seven}. A spatial-network perspective distinguishes connectivity from geographical embedding and examines how spatial constraints and connection costs influence network structure \cite{barthelemy2011,gastner2006}. Street-network analyses have also identified a structural distinction between a high-centrality (betweenness) backbone and lower-centrality loops providing local alternative paths \cite{kirkley2018}. Repeated mobility sustains connections between places, while these connections in turn influence subsequent movement and the location of activities \cite{hansen1959,louf2014congestion}. Through this reciprocal process, urban structures retain the imprint of earlier interactions and condition future development \cite{batty2013new,portugali2026}. Human settlement is itself unevenly distributed with elevation, with population concentrated at low altitudes \cite{CohenSmall1998}. Topography takes part in these spatial processes by shaping the possibilities and costs of connection \cite{GU2025105848,WANG2026107308}. Geographic obstacles have been linked to urban shape, growth and accessibility \cite{harari2020,wang2024}. Models in which settlements and transport networks co-evolve \cite{barthelemy2008modeling}, including over real landscapes \cite{aoki2022}, further show how geographical constraints can influence emergent spatial patterns through feedback between settlement and connectivity.
 
\begin{figure}[t]
    \centering
    \includegraphics[width=8.5cm]{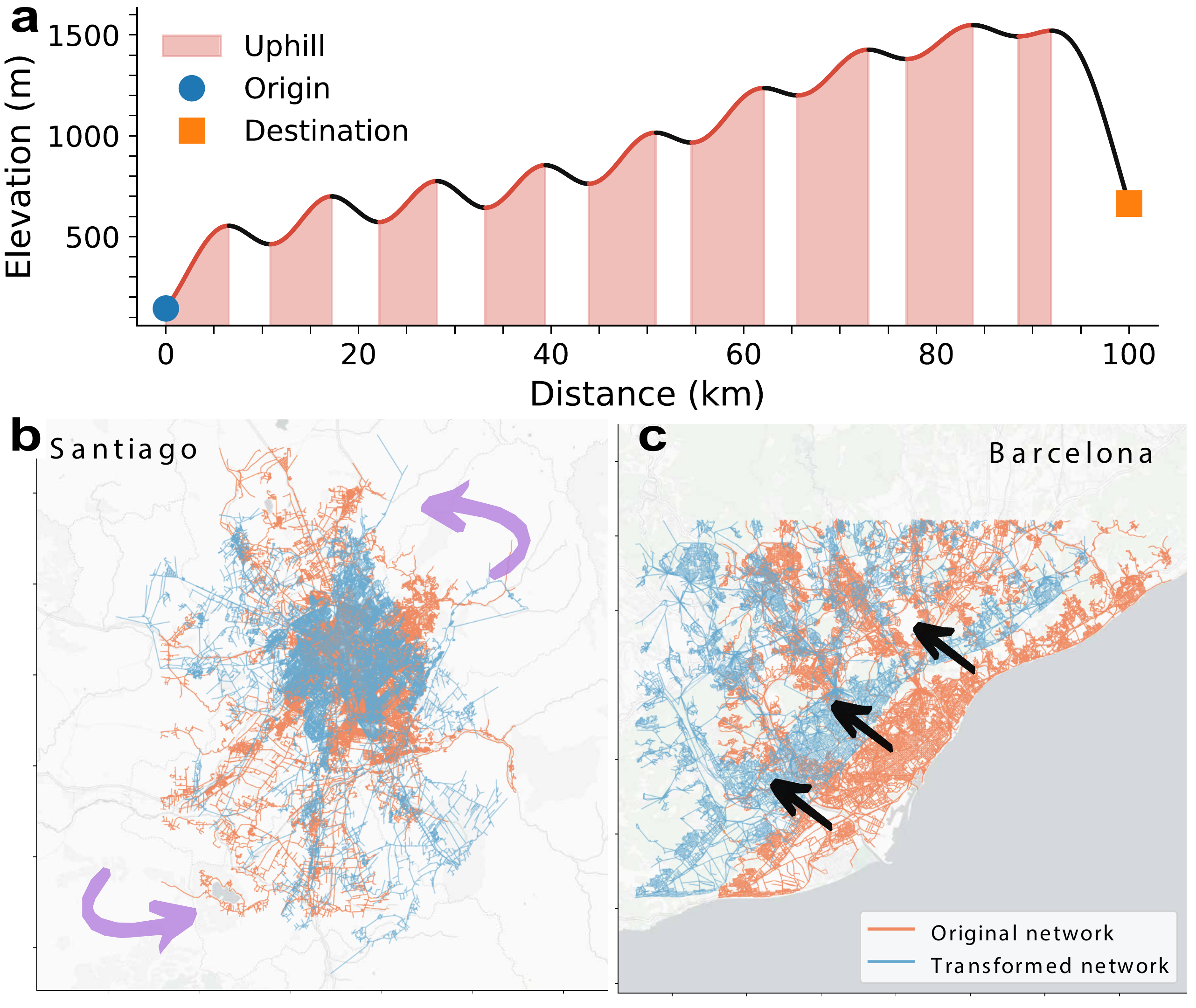}
    \caption{\textbf{Geometric transformations and topographic interpretation.}
    (\textbf{a}) Example elevation profile along a representative path: red segments indicate uphill portions contributing to gravitational work, while black segments correspond to downhill or flat portions (not contributing) to altitudinal energy. The total path length contributes to the horizontal energetic costs.
    (\textbf{b, c}) Spatial embedding of an urban road network under a representative transformation.
    The original configuration is shown in orange and the transformed configuration in blue, illustrating how pure geometric operations (translation/rotation pictured with the arrows) preserve topology while changing the alignment with the surrounding landscape.}
    \label{fig:fig1}
\end{figure}

Empirical research has established several dimensions of the relationship between urban configuration, mobility and physical geography. Travel behavior is associated with the density, diversity and design of the built environment, the spatial organization of activities, accessibility to destinations and street-network structure \cite{cervero1997travel,ewing2010,camagni2002urban}. Accessibility measures combine land use and transport to evaluate how easily destinations can be reached, typically expressing travel impedance through distance, time or generalized cost \cite{handy1997measuring,geurs2004}. Reducing the energy requirements of urban travel is a central aim of sustainable mobility policies \cite{banister2008sustainable}, and urban metabolism approaches treat energy flows as a constitutive dimension of urban functioning \cite{newman1999sustainability}. At the city scale, morphological and socioeconomic characteristics also provide complementary information about transport energy consumption \cite{lenechet2012urban}. Topography adds a further territorial dimension: terrain roughness and natural barriers are associated with differences in road network geometry and circuity \cite{GU2025105848,barthelemy2025}, while geographical fragmentation has been linked to urban compactness, spatial fragmentation and systematic differences in accessibility, including greater residential-employment separation \cite{WANG2026107308}. Terrain also affects the effort of non-motorized travel, as the energy cost of walking and running rises steeply on uphill slopes \cite{minetti2002energy}. Together, these studies connect urban configuration with mobility and its energetic requirements, while showing how physical geography relates to both spatial form and the connections through which urban activity is sustained.

This body of work leaves open whether existing urban configurations exhibit a generic energetic advantage in their relationship with the surrounding terrain. Routes of similar horizontal length can involve different cumulative elevation gains. The associated energy requirements recur whenever these connections are used, so their significance depends on both the routes available and the flows they carry. They therefore constitute a persistent condition under which connections between people, activities and places operate and evolve. This raises a question about urban spatial organization: do observed configurations exhibit systematically lower terrain-related energy requirements than alternative spatial configurations within the same topographic settings and under the same mobility demand?

To address this question, we define a comparative measure of movement energy that combines distance-related and altitudinal components to examine the relationship between urban configuration and topography. 
We infer origin-destination flows from geolocated records, assign routes from OpenStreetMap road networks and use digital elevation models to account for cumulative uphill movement. We separate the energy associated with each trip into a horizontal contribution, determined by route length, and an altitudinal contribution, determined by positive elevation gain. We then compare the observed configuration with translated, rotated and rescaled alternatives of the same network under identical mobility demand. This comparison isolates how the location, orientation and scale of an urban configuration relative to the terrain alter its movement energy requirements. Translations and rotations preserve network topology and internal distances while changing the position or orientation of the network relative to the terrain.
Across 56 cities, observed configurations consistently show lower movement energy requirements than most of the alternatives examined. This generic pattern reveals a measurable relationship between urban configuration and terrain that is compatible with cumulative adaptation among urban structure, mobility and topography. The comparisons identify a spatial regularity in existing configurations, although the analysis does not determine how this pattern emerged over time. By making this relationship measurable and comparable, the study provides empirical evidence on how movement requirements are associated with urban spatial organization. For urban planning, this introduces a measurable physical dimension that can inform the assessment of future urban growth and densification.

\begin{figure}[!t]
    \includegraphics[width=8.5cm]{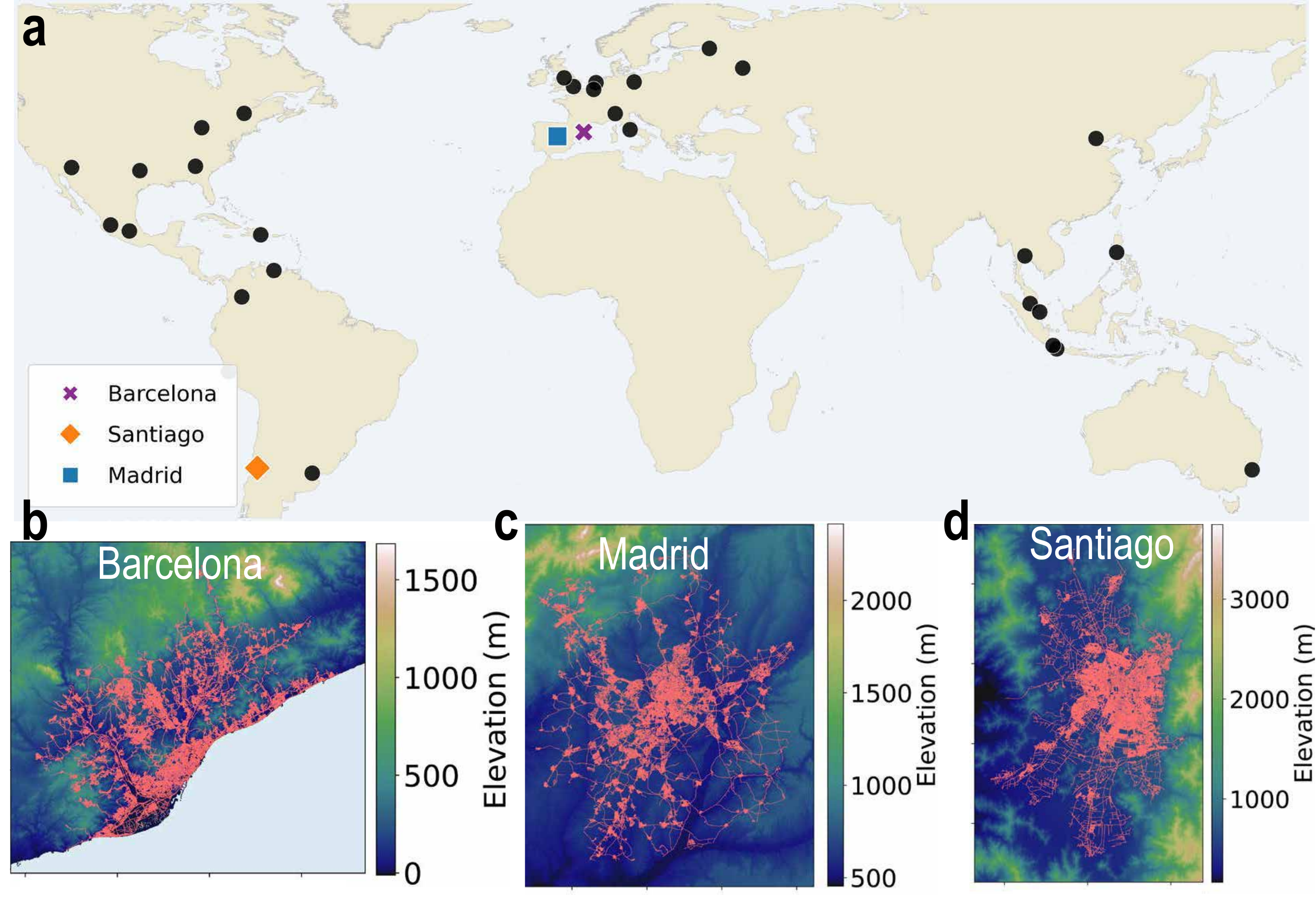}
    \caption{\textbf{Study areas and urban topography.}
    (\textbf{a}) Location of the analyzed cities.
    (\textbf{b--d}) Digital elevation models of Barcelona, Madrid and Santiago. Road networks are overlaid in light red.}
    \label{fig:fig2}
\end{figure}

\section*{Results}

\subsection*{Comparing urban configurations within their topographic settings} 

We examine the relationship between urban configuration and terrain by comparing each observed configuration with spatially transformed alternatives of the same road network. Keeping network connectivity and origin–destination demand fixed allows us to assess whether the observed configuration requires less movement energy for the represented mobility demand than the alternatives examined. This comparison isolates the energetic effect of how the urban network is positioned relative to the terrain.

For each trip represented in the empirical origin-destination data, we assign a trajectory on the road network and evaluate its energetic cost along the resulting three-dimensional path (see Methods for the underlying grid, mobility data, and route-assignment procedure). Throughout, we hold the energetic parameterization of a trip -- effectively that of private-vehicle travel -- fixed across every spatial configuration of a city we compare (see Methods for how $\lambda$ and $m$ are set). This choice isolates the effect of how the street network sits relative to the terrain from that of transport mode; it does not imply that the resulting mobility energy is inherently independent of mode, only that mode is held constant while spatial embedding varies.

For each assigned trajectory, the comparative measure combines a distance-related component, proportional to route length, and an altitudinal component, representing the work associated with cumulative uphill movement,
\begin{equation}\label{eq:tot_work}
    \begin{aligned}
    W_{\tau} & =  W_{\textrm{Hor}}(\tau) + W_{\textrm{Alt}}(\tau) =  \\
    & = \lambda \, L_\tau + m \, g \int_{0}^{L_\tau} \max\!\left(0, \frac{dh}{d\ell}\right) d\ell,
    \end{aligned}
\end{equation}
where $L_\tau$ is the length of trajectory $\tau$, $\lambda$ is the average energetic cost per unit distance, held fixed across the configurations compared (see Methods for its parameterization), $\ell$ is the curvilinear coordinate along the trajectory, $h(\ell)$ is the elevation at that point, $m$ is an effective transported mass, and $g$ is the gravitational acceleration. The integral only accumulates positive elevation gain, so downhill segments do not offset previous climbs (see Methods for the numerical implementation). Figure \ref{fig:fig1}{\bf a} illustrates this decomposition along a representative trajectory.

Summing over all $N$ trips $k=1,\dots,N$ recorded in a city gives the aggregate movement energy for a given configuration,
\begin{equation}
W_{\textrm{Tot}} = \sum_{k=1}^{N} W_{\tau_k} ,
\end{equation}
where $\tau_k$ denotes the trajectory of trip $k$.
We can then rotate and translate the street networks to explore the energetic costs of a different city position and shape (Fig.~\ref{fig:fig1}{\bf b,c}). We also tested stretching each city unevenly along its two main axes (anisotropic rescaling).

We have obtained Origin-Destination (OD) matrices for the FUAs of the cities shown in Figure~\ref{fig:fig2}{\bf a}. Figure~\ref{fig:fig2}{\bf b--d} depicts the bounding box covering the FUA of some cities together with the corresponding street networks extracted from OpenStreetMap. These representations combine the spatial domain of analysis with the underlying transport infrastructure, providing a consistent framework across different urban contexts. The overlay of the road network on the elevation field highlights how mobility pathways are embedded within the surrounding terrain.

Taken together, these maps display the key ingredients required for this study: the spatial extent of the urban system, the geometry of the street network, and the topography of the terrain within and around each city. This combination allows us to evaluate how mobility flows interact with elevation gradients, and to quantify the energetic implications of moving across heterogeneous landscapes.

\subsection*{Unequal distribution of altitudinal work}

\begin{figure}[!t]
    \centering
    \includegraphics[width=8.5cm]{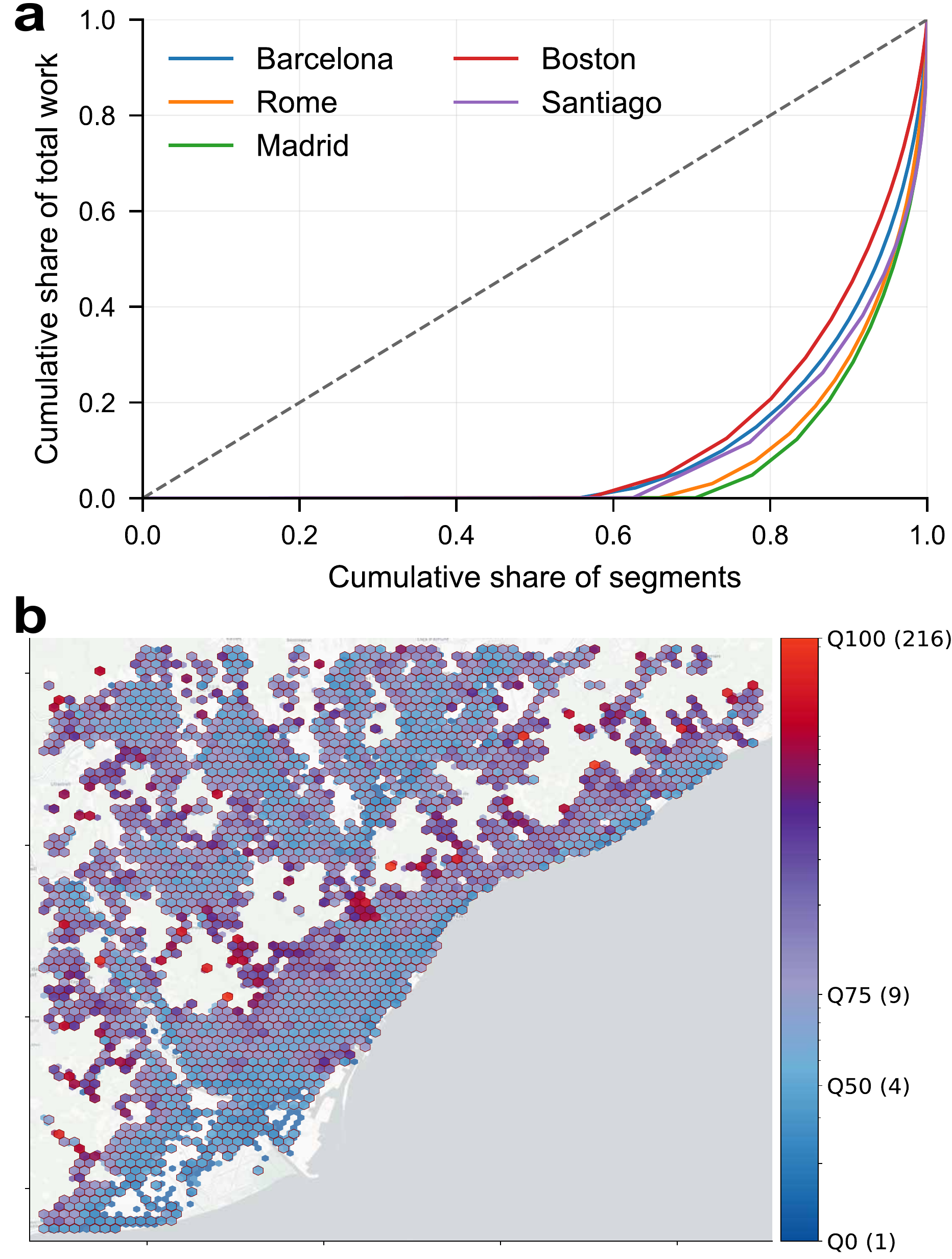}
    \caption{\textbf{Spatial concentration and inequality of altitudinal work.}
    (\textbf{a}) Lorenz curves of segment-level uphill altitudinal work for the selected cities.
    The dashed diagonal indicates perfect equality; deviations quantify concentration of energetic burden on a small fraction of links.
    (\textbf{b}) Example spatial map for Barcelona showing the distribution of altitudinal work aggregated over a hexagonal grid; colors denote quantile classes (Q0--Q100), highlighting localized high-work corridors.}
    \label{fig:fig3}
\end{figure}

We first examine how the altitudinal component of the comparative measure, $W_{\textrm{Alt}}$, is distributed across the road network. Figure~\ref{fig:fig3}{\bf a} shows that roughly $10\%$ of street network links account for more than $60\%$ of the aggregate altitudinal work (which yields a Gini index $\approx 0.8$), indicating a strong concentration of this component on a small part of the network. Figure~\ref{fig:fig3}{\bf b} shows Barcelona as a heatmap of $W_{\textrm{Alt}}$, using the quantiles of the altitudinal energetic cost accumulated by trajectories passing through each location. This map is purely a visualization device and uses a different spatial resolution from the grid used to define origin-destination cells in the rest of the analysis. Altitudinal work is a street-segment-level quantity, while aggregating it up to the origin-destination cells would blur its structure. For the sake of clarity, we have binned it onto a finer hexagonal grid with roughly $150$--$200\,\mathrm{m}$ sides, obtained by dividing the extent of the city into $200$ bins per side. Hexagonal cells are used to differentiate them from the original square cells of the OD flows.

High-work segments organize into coherent corridors where steep terrain coincides with the streets that carry the largest share of uphill flow. The energetic burden of a segment therefore depends jointly on two factors: how steep it is, and how much of the city's mobility demand is actually routed through it. A small subset of segments combines both conditions (steep grade and high flow) and together these form a backbone that channels most of the uphill mobility work in the city.

Because this concentration depends on which streets carry flow, it is sensitive to the geometric relationship between the network and the terrain. Translating or rotating the street network changes which segments intersect the steepest slopes and how flow is routed through them: even a modest shift can move a large share of trips onto, or away from, a handful of high-gradient corridors, producing disproportionate changes in the total energetic cost. This is precisely the mechanism behind the strong response to the geometric transformations examined below.

\subsection*{Urban configurations under energetic constraints}

\begin{figure}[!t]
    \includegraphics[width=8.5cm]{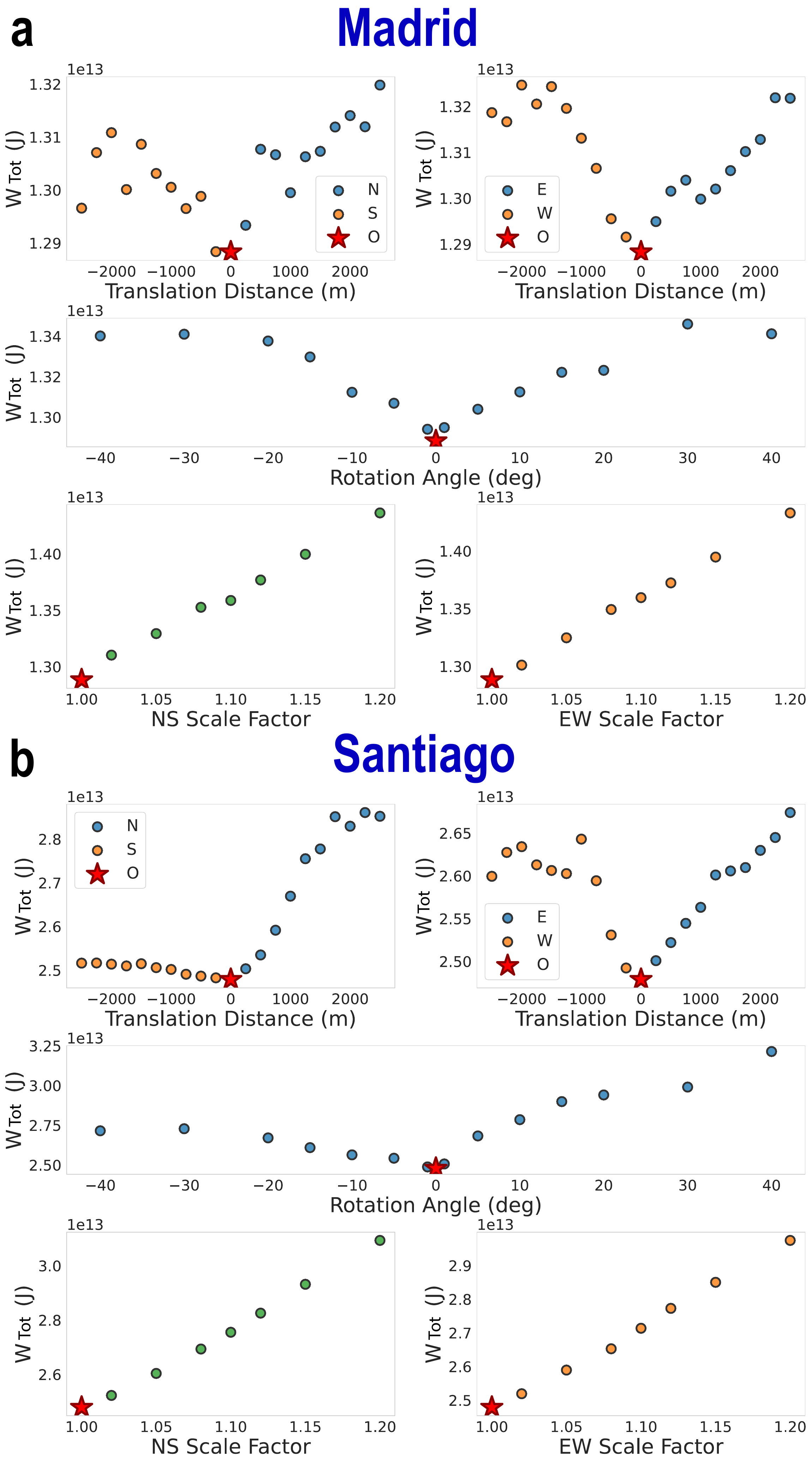}
    \caption{\textbf{Energetic sensitivity to geometric transformations of urban form.}
    Total mobility energy $W_{\mathrm{Tot}}$ (expressed in Joule) measured under controlled geometric transformations of the empirical urban configuration.
    (\textbf{a}) Madrid and (\textbf{b}) Santiago.
    For each city, the upper panels show North--South and East--West translations of the urban structure, the central panel shows rotations around the network centroid, and the lower panels show anisotropic scaling along the North--South and East--West directions.
    Colored points correspond to transformed configurations, while the red star marks the empirical city layout.
    In both cities, the original configuration consistently lies near the lowest energy expenditure observed across the tested transformations, suggesting that the observed spatial organization sits in a low-energy region of this transformation ensemble.}
    \label{fig:fig4}
\end{figure}

We then examine whether existing urban configurations correspond to lower energetic costs than alternative spatial arrangements within the same topographic domain and under the same mobility demand. As a first step, we focus on translations, rotations and rescalings of the cities. Systematic city transformations are displayed in Fig.~\ref{fig:fig4} for Madrid (Spain) and Santiago (Chile). For translations, we have moved the cities in steps of $250\,\mathrm{m}$ in directions north--south and east--west until $2.5\, \mathrm{km}$, while the rotations have been carried out using the geographical center of the street network as pivot point. Both types of transformations increase the mobility energy, having the original configuration as clear minimum (star). In Santiago the tallest mountains lie towards the east and north of the city and, therefore, translations in these directions consistently increase energetic costs showing the relevance of the altitudinal component. The rescaling transformations do not have a symmetric form, as stretching the city along different directions (north--south and east--west) changes both route lengths and the spatial relationship between the network and the terrain. As route lengths increase, the distance-related component grows and can dominate $W_{\text{Tot}}$.

\begin{figure}
\centering
\includegraphics[width=7cm]{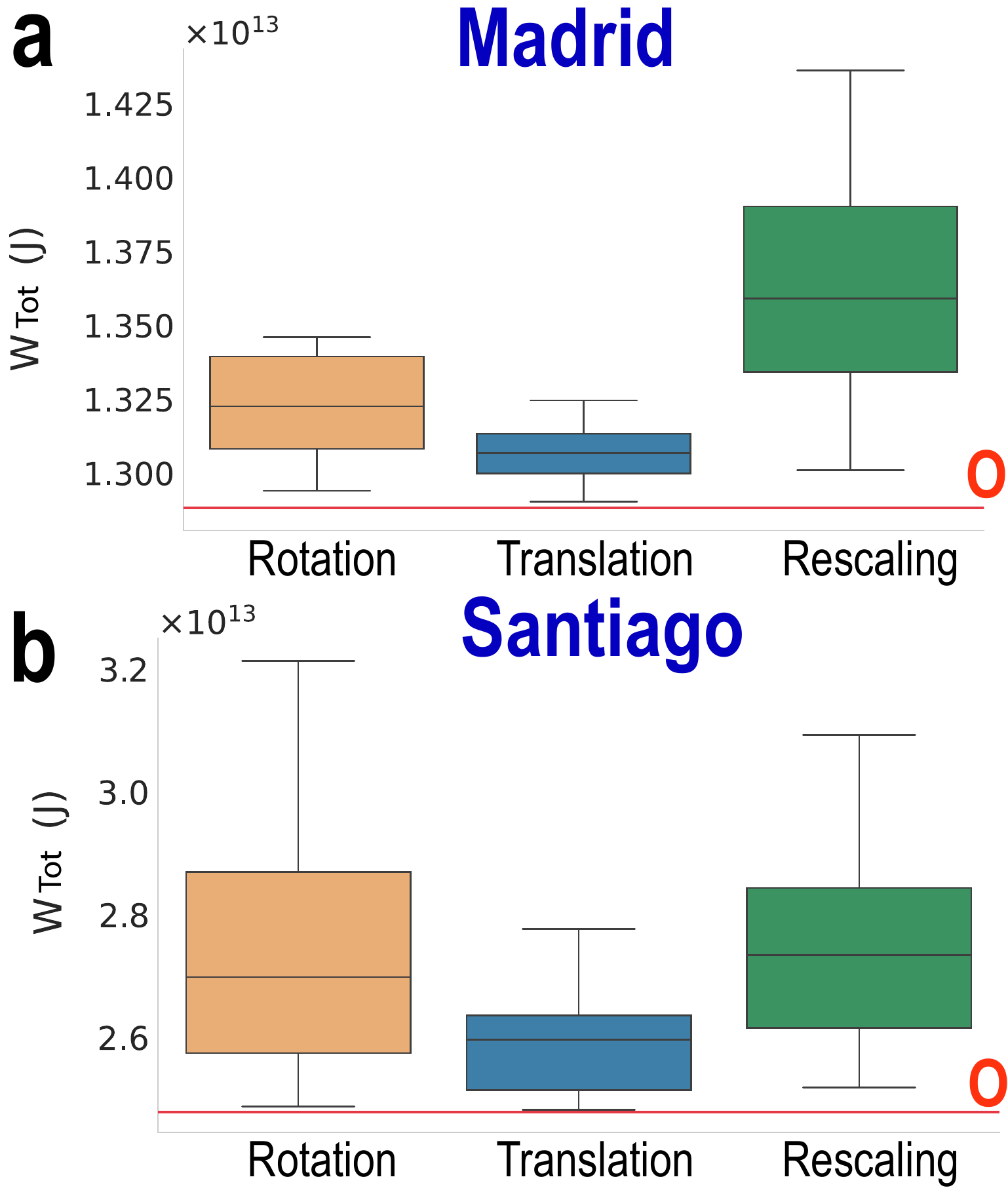}
\caption{\textbf{Energetic ranking under random geometric transformations.}
Total mobility energy $W_{\mathrm{Tot}}$ (expressed in Joule) computed for ensembles of randomly transformed urban configurations.
(\textbf{a}) Madrid and (\textbf{b}) Santiago.
The two panels summarize the corresponding energy distributions through box plots for each class of transformation. The horizontal red line, labeled `O', marks the total mobility energy of the empirical (observed) configuration.
In both cities, the empirical configuration systematically exhibits the lowest or near-lowest energetic cost, remaining below the bulk of transformed configurations.
These results indicate that the observed urban layouts are not arbitrary geometric arrangements, but configurations among the lowest-energy ones found within the tested transformation space.}
\label{fig:fig5}
\end{figure}

To ensure that this result is not limited to a few transformations, Fig.~\ref{fig:fig5} extends the analysis to broad ensembles including random displacements, rotations and rescaling. The empirical layout lies at the lower boundary of the distribution of $W_{\text{Tot}}$, while rotations, translations and rescalings generally exhibit higher values. Together, Figs.~\ref{fig:fig4} and \ref{fig:fig5} show that the observed embeddings remain near the lower boundary of the tested transformation ensemble, not only under the small, systematic transformations of Fig.~\ref{fig:fig4} but also across broad, randomized ones.

\subsection*{Exceptions to the low-energy pattern}

\begin{figure}[!t]
    \centering
    \includegraphics[width=9cm]{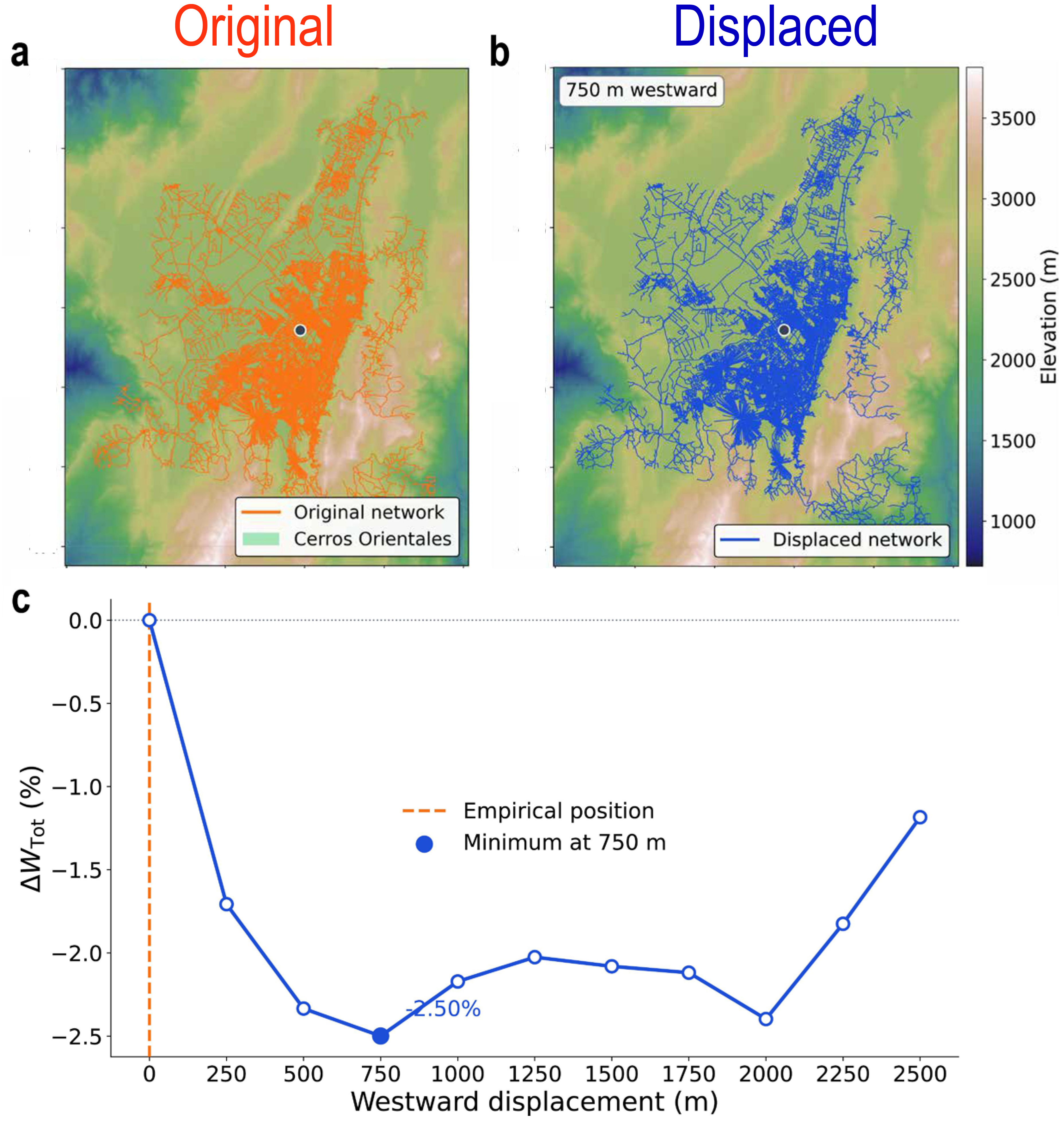}
        \caption{\textbf{Displaced configuration with lower mobility energy: Bogot\'a.}
        (\textbf{a}) Empirical road network overlaid on the local digital elevation model, with the Cerros Orientales protected reserve indicated to the east.
        (\textbf{b}) Lowest-energy displaced configuration: a westward translation of $750\,\mathrm{m}$, reducing total mobility energy by $2.5\%$.
        (\textbf{c}) Percentage change in total mobility energy relative to the observed configuration as a function of westward displacement. The vertical dashed line marks the empirical position; the curve reaches its minimum near $750\,\mathrm{m}$.}
    \label{fig:fig6}
\end{figure}

This analysis was performed in $56$ cities worldwide (Supplementary Table~1). In $26$ of them, located next to seas, lakes or the edge of the valid terrain domain, most transformations place part of the network outside that domain, so the comparison could be performed only partially (Supplementary Note~5). Among the remaining $30$ cities, the observed configuration is the lowest-energy one among the tested transformations in $26$ (Supplementary Notes~2 and 3 and Supplementary Figs~2--25). In the other four cases a transformed configuration achieves modestly reduced mobility energy: Bogot\'a ($-2.5\%$), Bandung ($-2.4\%$), Buenos Aires ($-0.7\%$) and Brussels ($-0.5\%$). In each case, the deviation is small  and a  geographic rationale can be identified, making these cases informative (see Supplementary Note 4 and Supplementary Figs 1, 26 and 27).

Bogot\'a provides the most compelling illustration (Fig.~\ref{fig:fig6}). Bogot\'a occupies the Bogot\'a savanna, a high-altitude plateau bordered to the east by the steep Cerros Orientales. The best-displaced configuration is a westward translation of $750\,\mathrm{m}$, which moves the road network slightly away from the Andes mountain-front elevation gradients and reduces cumulative altitudinal costs. This energetic advantage is physically consistent with the asymmetric topographic setting of the city. However, westward displacement is not a valid urban alternative. The city expanded to the west until the Bogot\'a river, whose corridor is associated with flood-prone areas, wetland restoration, flood-control works and inter-municipal planning involving Cota, Funza, Mosquera and Soacha \cite{bogota1, bogota2, bogota3}. The $-2.50$\% advantage should therefore be interpreted as a geometric low-energy scenario that is not feasible: the unconstrained energetic minimum falls partly on flooding areas and outside the hydrologically, ecologically and administratively accessible development space.

Geographic and regulatory constraints also provide possible explanations for the small departures observed in Bandung, Buenos Aires and Brussels (Supplementary Note 4). These cases highlight the distinction between geometrically generated alternatives and configurations that are feasible for urban development. Protected land, wetlands and existing infrastructure can restrict the set of feasible alternatives, so a lower movement energy cost need not correspond to a feasible urban alternative.

\subsection*{Application to densification scenarios}

\begin{figure}[!t]
    \centering
    \includegraphics[width=8.5cm]{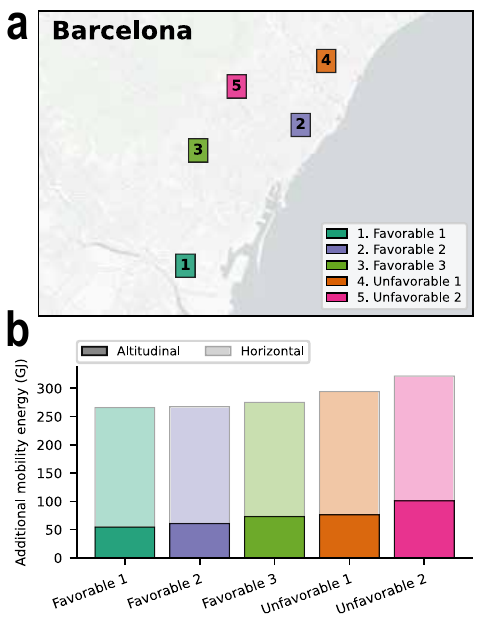}
    \caption{\textbf{Energetic evaluation of alternative densification scenarios in Barcelona.}
    (\textbf{a}) Spatial location of the five candidate grid cells (about $0.75\,\mathrm{km}\times0.75\,\mathrm{km}$) considered for urban densification, each receiving $\Delta P = 25{,}000$ new residents.
    Green and blue markers indicate energetically favorable configurations, while orange and magenta markers identify unfavorable ones.
    (\textbf{b}) Additional mobility energy generated by each scenario, decomposed into horizontal and altitudinal contributions.
    The total additional mobility energy is computed using gravity-model mobility demand (Eq.~(\ref{eq:gravity})), terrain-aware Dijkstra routing and the aggregation in Eq.~(\ref{eq:dens}), with route costs evaluated using Eq.~(\ref{eq:tot_work}).
    The selected scenarios exhibit different altitudinal contributions despite similar values of the distance-related component.}
    \label{fig:fig7}
\end{figure}

We illustrate the planning relevance of the framework by examining how alternative locations for additional population change the movement requirements within an existing urban configuration. Using Barcelona as a case study, we compare five densification scenarios under a common set of assumptions about the mobility generated by new residents.

Current population is taken from WorldPop 2020 and aggregated onto the same grid used in the mobility analysis. Each populated cell is then connected to the nearest node of the road network.

For each candidate location, we estimate the mobility demand generated by the new residents using a gravity model (see Methods section). The added population produces new trips, which are distributed across destination cells according to their population and their distance from the candidate cell. More populated and closer cells therefore attract larger flows, while longer-range interactions remain possible through an exponential distance-decay term with a characteristic scale $d_0\approx5.7\,\mathrm{km}$, calibrated on Barcelona's own empirical commuting flows (see Methods). This produces, for each candidate area, a synthetic OD demand describing the additional mobility generated by the densification scenario.

The generated trips are then routed on the real road network along the paths that minimize the terrain-following length $L_e$ (see Methods). The energetic cost of each route is evaluated using Eq.~(\ref{eq:tot_work}), combining the horizontal distance-dependent contribution and the uphill gravitational energetic cost accumulated along road segments. We then aggregate these route costs over all generated trips to obtain the total additional mobility energy associated with each candidate densification site.

The comparison between the selected candidate locations is shown in Fig.~\ref{fig:fig7}.
Panel~\textbf{a} reports the spatial distribution of five densification sites within the Barcelona FUA, selected among candidate cells whose horizontal mobility cost lies within $\pm5\%$ of the median.
Panel~\textbf{b} shows the corresponding additional mobility energy generated by adding $\Delta P = 25{,}000$ new residents, decomposed into horizontal and altitudinal components.

Despite similar values of the distance-related component, the total additional daily mobility energy ranges from about $266\,\mathrm{GJ}$ to nearly $322\,\mathrm{GJ}$.
The main differences emerge from the altitudinal component, which increases from approximately $55\,\mathrm{GJ}$ in the most favorable configuration to about $101\,\mathrm{GJ}$ in the least favorable one, nearly doubling the altitude-related energetic cost.
As a consequence, the least favorable location requires about $21\%$ more total mobility energy than the most favorable one, with the altitudinal component accounting for about $46$ of the $56\,\mathrm{GJ}$ that separate them.

These results show that energetic suitability cannot be inferred from geometric proximity alone.
Even locations with comparable travel-distance conditions may induce substantially different long-term mobility costs once terrain morphology and network structure are taken into account.
More generally, the example illustrates how the proposed framework can be used to examine the movement energy implications of alternative urban expansion or densification scenarios. In this perspective, terrain-related movement requirements provide an additional physical dimension for understanding how alternative growth locations relate to the existing urban configuration and topography.

\section*{Discussion}

Across the $56$ cities analyzed, observed urban configurations consistently show lower movement energy requirements than most of the translated, rotated and rescaled alternatives examined. This pattern occurs across topographically diverse settings, including cities without pronounced relief. The comparison identifies an empirical regularity in the relationship between urban configuration, mobility and terrain: for the mobility demand represented in the data, existing configurations tend to require less movement energy than the spatial alternatives tested. The extent to which this regularity holds beyond the cities and transformations examined remains to be established.
This recurrent pattern is compatible with cumulative adaptation among urban structure, mobility and terrain. Residence choices, infrastructure investment, planning decisions, institutions and repeated network use can all participate in the interactions through which urban configurations emerge and change. The observed configurations may therefore retain a spatial imprint of interactions accumulated over time, although the present analysis does not reconstruct their historical development. Topography forms part of these processes through the recurrent movements that connect people, activities and places, introducing physical requirements each time these connections are used. The energetic pattern identified here reveals one measurable physical dimension of their collective spatial outcome.

The first implication is that terrain contributes to city-scale spatial organization. Elevation gradients are not merely local nuisances: they accumulate along the trajectories that sustain urban functioning. This becomes explicit in the unequal distribution of segment cost. The Lorenz curves show that a small subset of links concentrates most uphill expenditure, meaning that energetic burden is concentrated on a sparse topographic backbone rather than being determined by average elevation alone. This also explains why the total cost responds so strongly to geometric transformations: even when topology is fixed, small translations and rotations change which streets carry the bulk of the flow, rerouting large fractions of demand across ridges, valleys, and slope discontinuities, and producing disproportionate changes in movement energy when high flows coincide with steep terrain.

The transformation analysis extends this finding beyond the immediate neighborhood of the empirical configuration. Across the systematic and randomized translated, rotated and rescaled alternatives of Figs.~\ref{fig:fig4} and \ref{fig:fig5}, the observed configurations repeatedly lie near the lower boundary of the tested transformation space. Conceptually, this disentangles three effects that are often conflated: topology determines connectivity, mobility demand determines how flows are distributed across the network,
and spatial embedding determines the energetic burden once terrain is introduced. More importantly, topology alone does not determine the energetic performance of an urban system. The same road network, supporting the same mobility demand, can sustain markedly different mobility costs depending solely on its position and orientation within the surrounding topography. We stress that this is a statement about the finite, physically motivated ensemble of transformations we tested, not a claim that the empirical configuration is a true optimum over the full, continuous space of physically accessible urban forms: a displaced city would itself adapt to its new local terrain over time, and our results say nothing about how that adapted form would compare to the configurations examined here.

For the four cities where a displaced configuration achieves lower $W_{\mathrm{Tot}}$, the discrepancy is small (below $3\%$) and arises from identifiable geographic constraints: protected forests, wetlands and flood-prone land, volcanic hazard zones and existing infrastructure. These constraints limit the accessible configuration space and explain why the empirical embedding does not coincide with the unconstrained geometric minimum. The observed urban form may already occupy the lowest-energy configuration available within the portion of the terrain that is historically, ecologically, or administratively accessible.

The densification scenarios illustrate why this relationship matters for urban planning. Adding population in different locations changes how new activity connects with the existing network and terrain, even where distance-related movement costs are comparable. The planning significance lies in making these physical consequences explicit when examining urban transformation. Expansion, densification and infrastructure investment establish connections whose movement requirements recur throughout their use, linking spatial decisions to the conditions that sustain urban activity over time.

Further work should test the robustness of the observed regularity across mobility datasets, transport modes and energetic parameterizations. Longitudinal analyses in which activities, mobility demand and network structure evolve together would also help establish how the relationship between configuration and terrain develops. These extensions would distinguish the recurrence of the spatial pattern from the particular assumptions used to measure it.

\section*{Methods}

\subsection*{Reconstructing the topographic landscape}

To reconstruct the physical landscape of each city, we used Shuttle Radar Topography Mission (SRTM) digital elevation model (DEM) data \cite{farr2007}. SRTM provides near-global elevation information at approximately $90\,\mathrm{m}$ resolution and offers a consistent topographic basis for comparing cities located in very different geographical settings. A comparison between the SRTM data and Copernicus DEM datasets is reported in Supplementary Note~7, showing only minor differences between the two products.

For each city, we downloaded the DEM tiles covering the corresponding functional urban area and its surrounding domain through OpenTopography's raster service \cite{SRTM-OpenTopography}. The elevation rasters were then reprojected from geographic coordinates to a local metric coordinate system. This was necessary because the quantities used in the analysis---road length, slope, trajectory distance, and elevation gain---must all be measured in compatible units. After this preprocessing step, the DEM defines a continuous elevation field $h(x,y)$ over which the road network and all its transformed versions can be evaluated.

\subsection*{Urban systems and spatial domains}

We applied the analysis to a set of $56$ cities chosen from different regions (see Supplementary Table~1). These cities combine the availability of empirical mobility data with diverse topographic contexts, from relatively flat metropolitan regions to urban areas constrained by hills, valleys, and mountain ranges. City selection was constrained primarily by the availability of geolocated mobility data meeting the filtering criteria described below; as a result, African cities are not currently represented in our sample. We regard this as a limitation of data availability rather than of the framework itself, and we intend to extend the analysis to additional regions, including Africa, as comparable data become available.

For each case, the spatial extent of the analysis was defined using the Functional Urban Area (FUA) boundary \cite{fua}. The FUA provides a standardized, functionally motivated delineation of the metropolitan system, capturing the full extent of commuting zones and recurrent mobility flows rather than the administrative boundary of the core municipality. This choice is critical: using a smaller administrative boundary would restrict the set of tested configurations and alter the outcome of the energetic comparison (see Supplementary Note~6 for a detailed comparison). All spatial layers were projected into the same metric coordinate system before being combined.

\subsection*{Empirical mobility demand}

Mobility demand was represented by origin-destination matrices between regular urban cells. Each functional urban area was divided into square cells of side length $1\,\mathrm{km}$ in the Web Mercator projection (EPSG:3857), whose ground size decreases with latitude (about $0.75\,\mathrm{km}$ at Barcelona), and observed movements were aggregated between these cells. The resulting quantity $F_{OD}$ measures the number of trips from an origin cell $O$ to a destination cell $D$.

The empirical flows were derived from geolocated Twitter/X data taken from 2015 to 2022. To focus on repeated urban mobility rather than occasional visits, we retained only users with sustained presence in the same city over at least two months. This filtering step is intended to capture routine movements associated with resident populations. Beijing and Shanghai do not satisfy this criterion in practice: Twitter/X is officially blocked in mainland China, so the residual geolocated activity we observe in these two cities is dominated by foreign visitors rather than by the resident population, a bias we could not correct with the data available to us. We therefore removed both cities from the dataset. Previous studies have shown that individual mobility is highly regular and predictable at aggregate scale \cite{gonzalez2008understanding, song2010limits}, making OD matrices a suitable representation of recurrent urban demand \cite{bassolas2019hierarchical}.

\subsection*{Road networks and route assignment}

The urban road system was extracted from OpenStreetMap using OSMnx \cite{boeing2017,boeing2025modeling}. Each city was represented as a directed graph $G=(V,E)$, where nodes $V$ correspond to intersections or road endpoints, and directed edges $E$ represent road segments.

For each road segment $e\in E$, we first computed its planar metric length $L_e^{2D}$. After sampling the corresponding elevation profile from the DEM, we assigned to the same segment a three-dimensional terrain-following length
\begin{equation}\label{eq:L3d}
    L_e = \sum_{k \in e}\sqrt{\Delta x_k^2+\Delta y_k^2+\Delta h_k^2},
\end{equation}
where $\Delta x_k$ and $\Delta y_k$ are metric horizontal increments and $\Delta h_k$ is the corresponding elevation difference along the segment.

For each OD pair, origin and destination points were sampled uniformly within their corresponding grid cells and then mapped to the nearest road-network nodes. The assigned trajectory was obtained with Dijkstra's algorithm \cite{dijkstra1959} by minimizing the cumulative three-dimensional path length,
\begin{equation*}
\tau_{OD} = \arg\min_{\tau:O\rightarrow D} \sum_{e\in \tau} L_e.
\end{equation*}
Routes were assigned by minimizing terrain-following path length rather than the energetic cost itself. This avoids biasing the route choice directly toward the quantity later used to evaluate mobility energy. Route assignment does not account for congestion, so assigned paths approximate free-flow conditions rather than actual travel-time-minimizing behavior under load.

\subsection*{Energetic decomposition of mobility}

Once a trajectory has been assigned to the road network, its energetic cost can be decomposed into two contributions:
\[
W_\tau = W_{\mathrm{Hor}}(\tau)+W_{\mathrm{Alt}}(\tau).
\]
The horizontal term is modeled as proportional to path length,
\[
W_{\mathrm{Hor}}(\tau)=\lambda L_\tau,
\]
where $\lambda=\eta E_l c=585\,\mathrm{J/m}$ is an effective energetic cost per unit distance combining fuel energy content $E_l=36\,\mathrm{MJ/L}$, fuel consumption $c=6.5\,\mathrm{L}/100\,\mathrm{km}$ (mid-range of $5$--$8\,\mathrm{L}/100\,\mathrm{km}$) and an effective conversion efficiency $\eta=0.25$. To assign concrete values to $\lambda$ and to the effective transported mass $m$ (see below), we parameterize trips as private-vehicle travel. This is a simplifying assumption: it fixes mode-specific numerical values for these parameters, held constant across every spatial configuration of a given city, rather than resolving how the energetic cost would change under a different mode split. Holding this mode-specific parameterization fixed provides a controlled comparison between the empirical configuration and its geometric alternatives.

The altitudinal contribution is the gravitational work associated with uphill motion,
\[
W_{\mathrm{Alt}}(\tau) = m \, g \int_{0}^{L_\tau} \max\!\left(0, \frac{dh}{d\ell}\right) d\ell,
\]
where $m=1200\,\mathrm{kg}$ is an effective transported mass, $g$ is gravitational acceleration, $h$ is elevation, and $\ell$ is the curvilinear coordinate. Only uphill portions contribute; downhill segments are not allowed to compensate previous climbs.

In the numerical implementation, each road segment $e$ is sampled every $\Delta s=10\,\mathrm{m}$. The positive elevation gain accumulated along the edge is
\[
\Delta h_e^{+} = \sum_{i=0}^{N_e-2} \max(0,h_{i+1}-h_i),
\]
where $N_e$ is the number of sampling points along edge $e$ and $h_i$ the elevation at the $i$-th point, giving the altitudinal edge work $W_{\mathrm{Alt}}(e)=m g \Delta h_e^{+}$ and the complete edge cost $W_e = \lambda L_e + m g \Delta h_e^{+}$.

The city-wide mobility energy is obtained by weighting every path by the corresponding empirical OD flow:
\[
W_{\mathrm{Tot}} = \sum_{(O,D)}F_{OD}W_{OD}.
\]

\subsection*{City replicas}

Each transformed configuration is obtained through
\[
\mathbf{x}'=\mathbf{A}\mathbf{x}+\mathbf{b},
\]
where $\mathbf{A}$ encodes rotation or rescaling, and $\mathbf{b}$ encodes translation. Translations ($\mathbf{A}=\mathbf{I}$) shift the entire network rigidly, preserving internal geometry while exposing it to a different portion of the terrain. Rotations around the network centroid modify the alignment between the street system and the dominant slope directions while preserving all internal planar distances. Rescalings around the centroid rescale the city along the north--south or east--west axes.

For every transformed city, the DEM was sampled again along all road segments and $W_{\mathrm{Tot}}$ was recomputed from the beginning. Transformations that placed any node of the network outside the valid terrestrial domain---over the sea, lake shores, or outside the valid DEM extent---were discarded. The remaining configurations define the set of counterfactual alternatives against which the empirical city is compared.

Transformation increments of $250\,\mathrm{m}$ in each of the four cardinal directions (up to a maximum displacement of $2{,}500\,\mathrm{m}$), rotation angles of $\pm 1^\circ$ to $\pm 40^\circ$, and anisotropic scaling factors from $1.02$ to $1.20$, i.e. expansions of the built area only, were tested. Contractions were excluded from this ensemble because shrinking the network compresses its nodes and edges into a footprint smaller than that of the empirical city, below the spacing imposed by its existing blocks and buildings. Any associated reduction in mobility energy could therefore arise trivially from shorter paths rather than from a more favorable alignment between the network and the terrain. We evaluated this interpretation in a separate sensitivity analysis for Madrid and Santiago, using contraction factors reciprocal to the expansion factors considered in the main analysis. In both cities, mobility energy generally decreased as the network was contracted, supporting the interpretation that the apparent energetic benefit is primarily a geometric compression effect rather than a terrain-alignment effect (Supplementary Note~8). We therefore did not include contractions in the main transformation ensemble. Nodes of the original road network that fall inside lake or coastal polygons (e.g., nodes on bridges) are excluded from the land-validity check, so that only nodes displaced from land to water by the transformation are counted as infeasible.

\subsection*{Energetic evaluation of densification scenarios}

The same framework was used to examine hypothetical densification scenarios. We first aggregated current population from the WorldPop 2020 unconstrained gridded population datasets \cite{worldpop} onto the same grid. The mobility demand generated by new residents was estimated with a gravity model \cite{wilson1971}:
\begin{equation}\label{eq:gravity}
    F_{OD} = N_{\mathrm{trips}}\frac{P_D^{\alpha}\exp(-d_{OD}/d_0)}{\sum_{K\neq O}P_K^{\alpha}\exp(-d_{OK}/d_0)},
\end{equation}
with $\alpha=1$ and $N_{\mathrm{trips}}=\Delta P$, corresponding to one daily round trip per new resident (e.g.\ home--work and work--home), each including the outbound and the return path (Eq.~(\ref{eq:dens})). Rather than fixing $d_0$ a priori, we calibrated it on Barcelona's own empirical working-day origin-destination flows, fitting the production-constrained model of Eq.~(\ref{eq:gravity}) (with $\alpha=1$ fixed) by Poisson pseudo-maximum-likelihood with origin fixed effects\cite{silva2006}. This is equivalent to maximizing the multinomial likelihood of Eq.~(\ref{eq:gravity}) and accounts for cell pairs without recorded trips through the normalization over all destinations. Distances are measured between cell centroids in the local UTM projection. The fit uses $151{,}733$ origin-destination pairs with recorded trips from $2{,}577$ origin cells ($900{,}538$ trips) and gives $d_0=5.7\pm0.2\,\mathrm{km}$ (origin-clustered robust standard error).

The total additional mobility energy induced by densifying cell $O$ is
\begin{equation}\label{eq:dens}
    W_{\mathrm{Tot,dens}}^{(O)} = \sum_{D\neq O} F_{OD} \left(W_{OD}+W_{DO}\right),
\end{equation}
decomposable as $W_{\mathrm{Tot,dens}}^{(O)} = W_{\mathrm{Hor,dens}}^{(O)} + W_{\mathrm{Alt,dens}}^{(O)}$.

For the densification scenarios, every road segment was projected to the local UTM coordinate system and sampled every $10\,\mathrm{m}$, with elevations obtained by bilinear interpolation of the DEM reprojected to the same system; trips were routed between the nearest network nodes of origin and destination cells by minimizing the terrain-following length $L_e$ (Eq.~(\ref{eq:L3d})). We evaluated $90$ candidate cells, chosen among the most populated central and peripheral cells of the FUA. Among the candidates whose horizontal component lies within $\pm5\%$ of the median, we selected the three with the lowest and the two with the highest altitudinal component, imposing a minimum separation of $3.75\,\mathrm{km}$ between selected cells. The comparison therefore isolates the role of terrain morphology at comparable distance-related costs, and the reported range corresponds to the extremes of this candidate set. \\

\section*{Data Availability Statement}

OpenStreetMap road network data can be accessed via OSMnx (\url{https://osmnx.readthedocs.io/}). SRTM digital elevation model data are publicly available from OpenTopography (\url{https://doi.org/10.5069/G9445JDF}). Map visualizations use CARTO Positron basemap tiles \cite{carto_positron}, with data by OpenStreetMap under ODbL. Aggregated origin-destination matrices generated and analyzed during the current study are available from the corresponding authors upon reasonable request; the underlying geolocated records were obtained under a data sharing agreement and cannot be shared.

\section*{Code Availability Statement}

Codes used to perform the analysis are available at
(\url{https://github.com/federicobellisardi/topolity})

\section*{Acknowledgements}

F.B., A.R.V. and J.J.R. acknowledge support by the Spanish State Research Agency
(MICIU/AEI/10.13039/501100011033) and FEDER (UE) under project COSASTI (PID2024-157493NB-C21 \& PID2024-157493NB-C22), and the Mar{\'\i}a de Maeztu project CEX2021-001164-M. A.R.V. acknowledges support as well from the ``Departamento de Ciencia, Universidad y Sociedad del Conocimiento'' of the Government of Aragon, Spain (S04\_23R). 

\section*{Author Contributions}
F.B., A.R.V., E.C.S. and J.J.R. conceived and designed the study. F.B. collected, pre-processed and analyzed the data. F.B. performed the simulations. F.B., A.R.V., E.C.S. and J.J.R. discussed the results. All the authors wrote and approved the paper.

\section*{Competing Interests}
The authors declare no competing interests.

\bibliographystyle{naturemag}
\bibliography{references}

\clearpage
\onecolumngrid
\setcounter{figure}{0}
\setcounter{table}{0}
\renewcommand{\figurename}{Supplementary Fig.}
\renewcommand{\tablename}{Supplementary Table}
\begin{center}
{\Large\bfseries Supplementary Information for}\\[8pt]
{\large Mobility energetic balance frames spatial configuration of cities}\\[6pt]
{\normalsize Federico Bellisardi$^1$, Enrique Cano Su\~n\'en$^2$, Ana Ruiz-Varona$^3$ and Jos\'e J.\ Ramasco$^1$}\\ \vspace{.8cm}
{\small $^1$Instituto de F\'{\i}sica Interdisciplinar y Sistemas Complejos IFISC (CSIC-UIB), Campus UIB, 07122 Palma de Mallorca, Spain \\$^2$Human Openware Research Lab (HOWLab), Engineering Research Institute of Aragon (I3A), University of Zaragoza (UZ), 50018 Zaragoza, Spain \\ $^3$Arquitecturas Open Source (AOS) Research Group, School of Architecture and Technology, Universidad San Jorge, 50830 Zaragoza, Spain }
\end{center}

\vspace{6pt}\hrule\vspace{10pt}

\section*{Supplementary Note 1: Overview of city-level results}

This supplementary document reports the full city-by-city validation of the hypothesis that the empirical urban configuration occupies a low-energy position within the tested set of geometric transformations. We evaluated 56 urban systems: in 26 cases the empirical embedding is the minimum-energy configuration within the tested transformation set (Supplementary Note~3); in 4 cases a displaced configuration achieves modestly lower energy, with a geographic rationale in each case (Supplementary Note~4); and 26 cities are constrained by coastlines, lakes or the limits of the data domain, which render most alternative configurations infeasible (Supplementary Note~5). Results by city are summarized in Supplementary Table~1.

Supplementary Fig.~\ref{fig:supp:counteroverview} summarizes the magnitude of the energy reduction in the 4 cases where a displaced configuration achieves lower $W_{\mathrm{Tot}}$.

Displacement configurations with lower $W_{\mathrm{Tot}}$ should be interpreted cautiously. A lower-energy transformed embedding in a partially constrained search space does not imply that the observed city is globally suboptimal. These cases may reflect territorial constraints, historical path dependence, protected areas, waterfront geometries, or infrastructure barriers.

\begin{figure}[ht]
\centering
\includegraphics[width=0.82\textwidth]{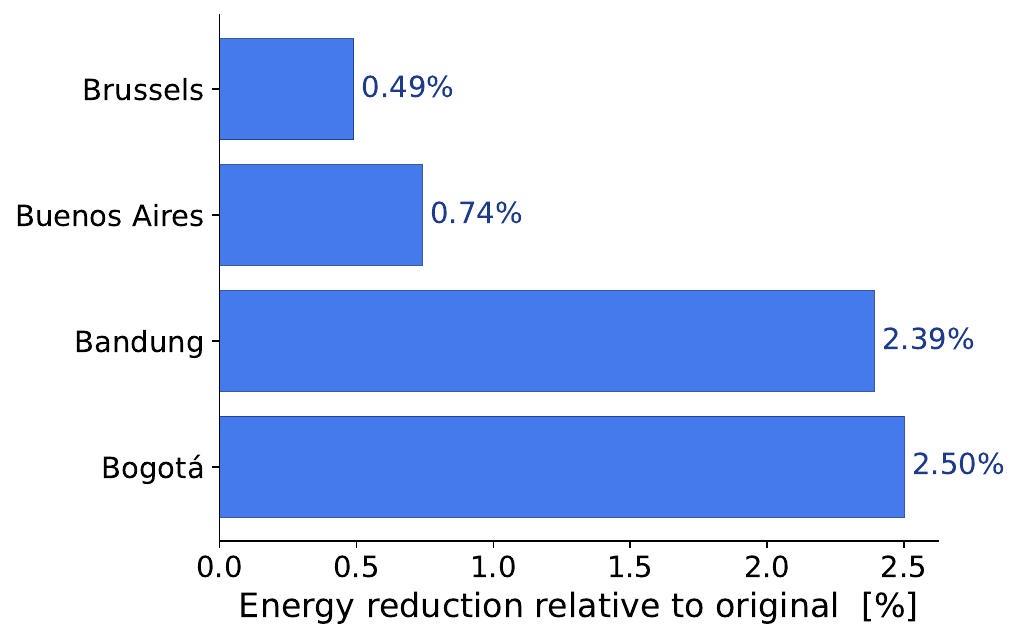}
\caption{\label{fig:supp:counteroverview}\textbf{Magnitude of the energy reduction achieved by the best displaced configuration in the 4 cities where such a configuration exists.} Bars show $|100\times(W_{\min}/W_{\mathrm{orig}}-1)|$, sorted by magnitude. All deviations are below $3\%$, indicating that the empirical embedding remains close in energy to the lowest-energy tested alternative.}
\end{figure}

\section*{Supplementary Note 2: City-level results summary}

Supplementary Table~1 reports the full city-level results. The following figures summarize the distribution of energy deviations across cities and the 4 cities where a displaced configuration achieves lower mobility energy.

\begin{longtable}{@{}ll@{}}
\caption{\label{tab:supp:summary}\textbf{City-level energetic validation summary.} Cities grouped by category: OK (Original), Exceptions (displaced lower), and Constrained. For OK cities the `Result' column simply reports `Original'.}\\
\toprule
\multicolumn{1}{@{}l}{\textbf{OK (Original)}} & \\
\midrule
City & Result \\
\midrule
\endfirsthead
\toprule
\multicolumn{1}{@{}l}{\textbf{}} & \\
\midrule
City & Result \\
\midrule
\endhead
Bangkok & Original \\
Barcelona & Original \\
Berlin & Original \\
Caracas & Original \\
Dallas & Original \\
Guadalajara & Original \\
Jakarta & Original \\
Kuala Lumpur & Original \\
Lima & Original \\
London & Original \\
Madrid & Original \\
Manchester & Original \\
Manila & Original \\
Mexico City & Original \\
Milan & Original \\
Montreal & Original \\
Moscow & Original \\
Paris & Original \\
Phoenix & Original \\
Rome & Original \\
Saint Petersburg & Original \\
Santiago & Original \\
Santo Domingo & Original \\
Singapore & Original \\
S\~ao Paulo & Original \\
Sydney & Original \\
\midrule
\multicolumn{1}{@{}l}{\textbf{Exceptions (Displaced lower)}} & \\
\midrule
City & Result \\
Bogot\'a & Displaced lower: Translation 750 W, $\Delta=-2.50\%$ \\
Bandung & Displaced lower: Translation 2500 S, $\Delta=-2.39\%$ \\
Buenos Aires & Displaced lower: Translation 1750 S, $\Delta=-0.74\%$ \\
Brussels & Displaced lower: Translation 2250 N, $\Delta=-0.49\%$ \\
\midrule
\multicolumn{1}{@{}l}{\textbf{Constrained}} & \\
\midrule
City & Result \\
Amsterdam & Constrained \\
Atlanta & Constrained \\
Boston & Constrained \\
Chicago & Constrained \\
Detroit & Constrained \\
Dublin & Constrained \\
Hong Kong & Constrained \\
Houston & Constrained \\
Istanbul & Constrained \\
Lisbon & Constrained \\
Los Angeles & Constrained \\
Miami & Constrained \\
Nagoya & Constrained \\
New York & Constrained \\
Osaka & Constrained \\
Philadelphia & Constrained \\
Rio de Janeiro & Constrained \\
San Diego & Constrained \\
San Francisco & Constrained \\
Seoul & Constrained \\
Stockholm & Constrained \\
Taipei & Constrained \\
Toronto & Constrained \\
Vancouver & Constrained \\
Washington D.C. & Constrained \\
Tokyo & Constrained \\
\bottomrule
\end{longtable}

\paragraph{Table notes.}
The ``Result'' column indicates whether the empirical embedding is the energetic \emph{Original} (minimum energy among all tested configurations), whether a \emph{Displaced lower} configuration (geometrically equivalent but lower energy) was found, or whether the city is \emph{Constrained} (most transformations are rejected and the remaining ones do not allow a meaningful comparison; see Supplementary Notes~3 and 5). For the displaced cities, the result entry also reports the best transformation family (translation, rotation, or scaling), the displacement direction and magnitude, and the relative energy deviation $\Delta = 100\times(W_{\min}/W_{\mathrm{orig}}-1)$ (negative values indicate the displaced embedding has lower energy).

\clearpage
\section*{Supplementary Note 3: Cities consistent with the low-energy hypothesis}
\label{sec1}

The 26 cities listed below satisfy the condition that $W_{\mathrm{Tot}}$ of the empirical configuration is lower than or equal to $W_{\mathrm{Tot}}$ of every tested geometric alternative derived from the same street network: Bangkok, Barcelona, Berlin, Caracas, Dallas, Guadalajara, Jakarta, Kuala Lumpur, Lima, London, Madrid, Manchester, Manila, Mexico City, Milan, Montreal, Moscow, Paris, Phoenix, Rome, Saint Petersburg, Santiago, Santo Domingo, S\~ao Paulo, Singapore, Sydney.

The number of admissible alternatives differs widely across these cities (Supplementary Fig.~\ref{fig:supp:landcounts}). In Bangkok, Berlin, Manila and Sydney a single transformation remains admissible, and in Montreal and London only three and four, respectively. For these six cities the comparison is therefore limited, and their classification as `Original' should be read accordingly.

For each city, the energetic sensitivity profile below displays how $W_{\mathrm{Tot}}$ varies as a function of the transformation parameter across the admissible embedding space. The empirical configuration is marked with a red star at the origin. Profiles are shown only for transformation families that produced at least one valid alternative embedding; families with no admissible configuration are omitted. Madrid and Santiago are the main showcase cities discussed in the main text (Fig.~4); their profiles are shown there and are not repeated below. The profiles of the remaining 24 cities are shown below for completeness.

\subsection*{Per-city energetic sensitivity profiles}

\begin{figure*}[ht]
\centering
\includegraphics[width=0.96\textwidth]{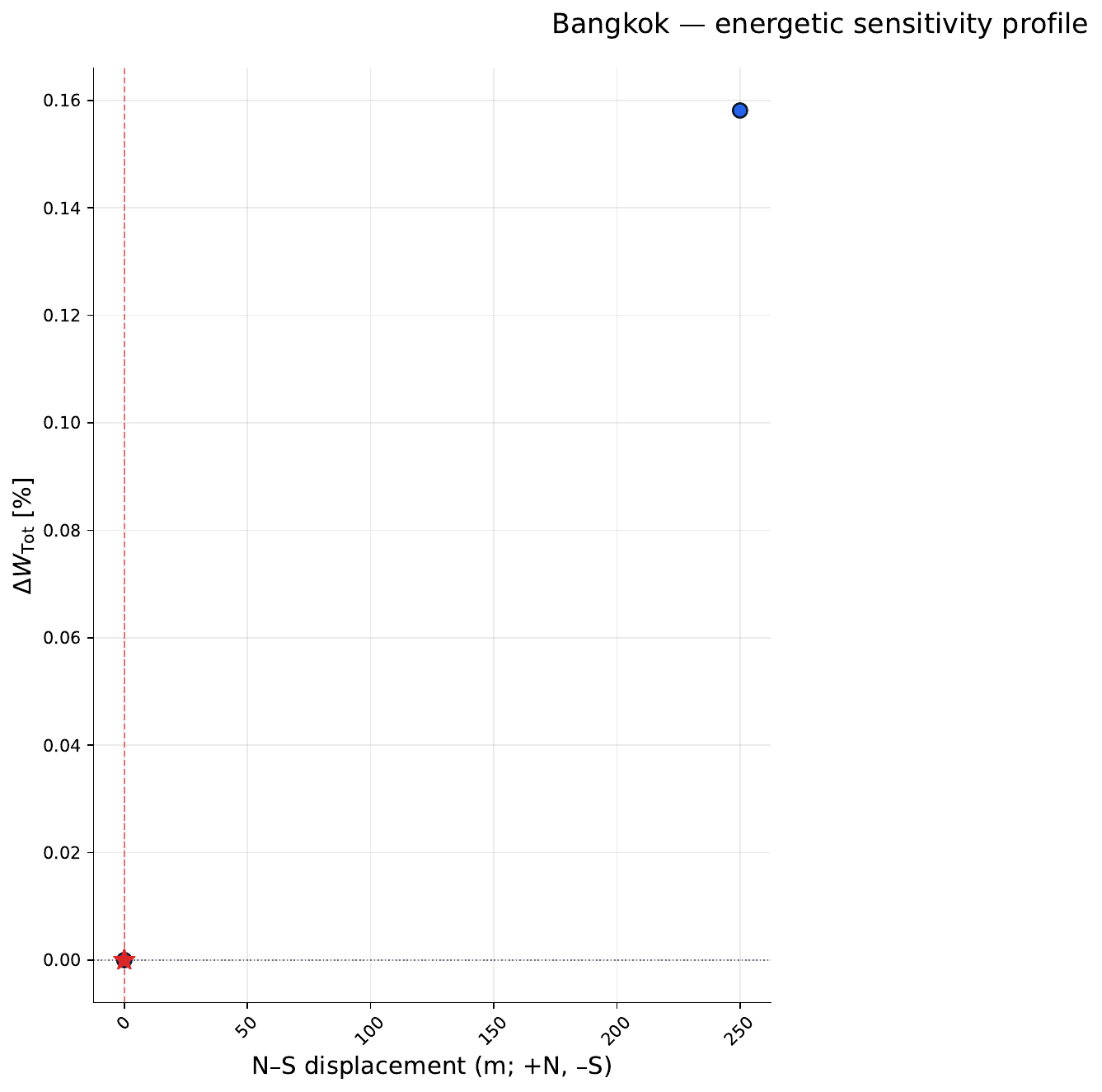}
\caption{\textbf{Bangkok — energetic sensitivity profile.}}
\end{figure*}

\begin{figure*}[ht]
\centering
\includegraphics[width=0.96\textwidth]{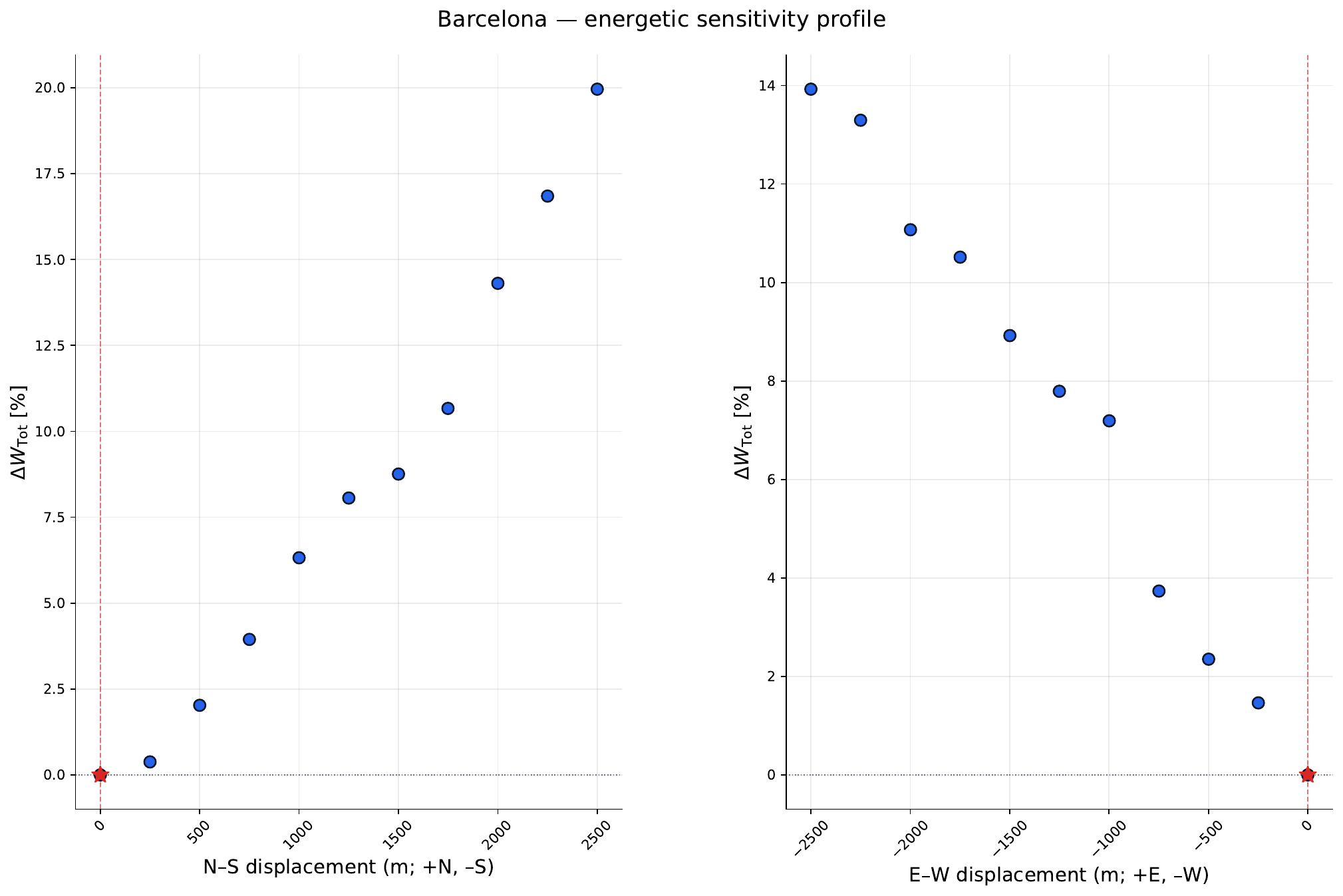}
\caption{\textbf{Barcelona — energetic sensitivity profile.}}
\end{figure*}

\begin{figure*}[ht]
\centering
\includegraphics[width=0.96\textwidth]{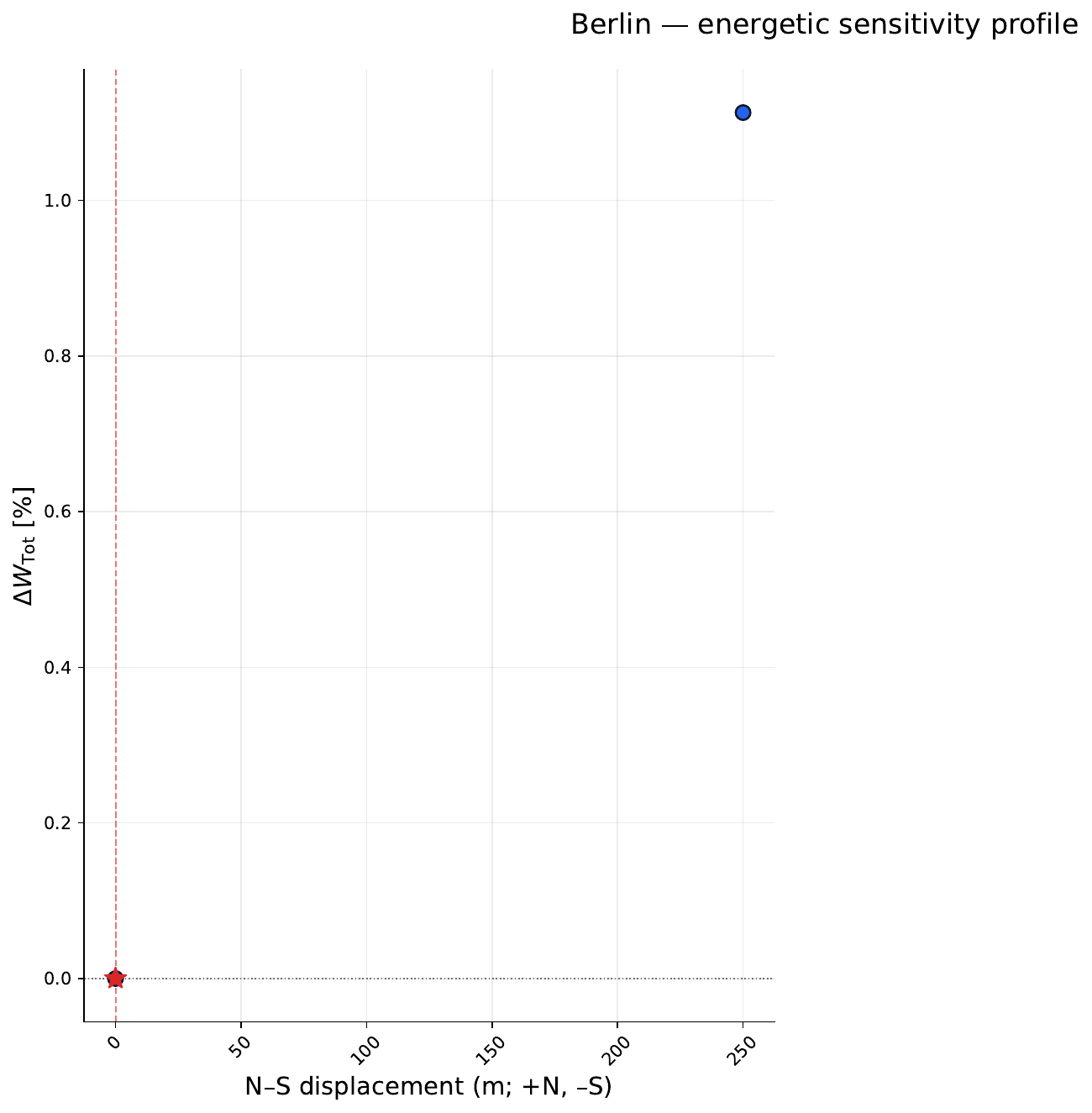}
\caption{\textbf{Berlin — energetic sensitivity profile.}}
\end{figure*}

\begin{figure*}[ht]
\centering
\includegraphics[width=0.96\textwidth]{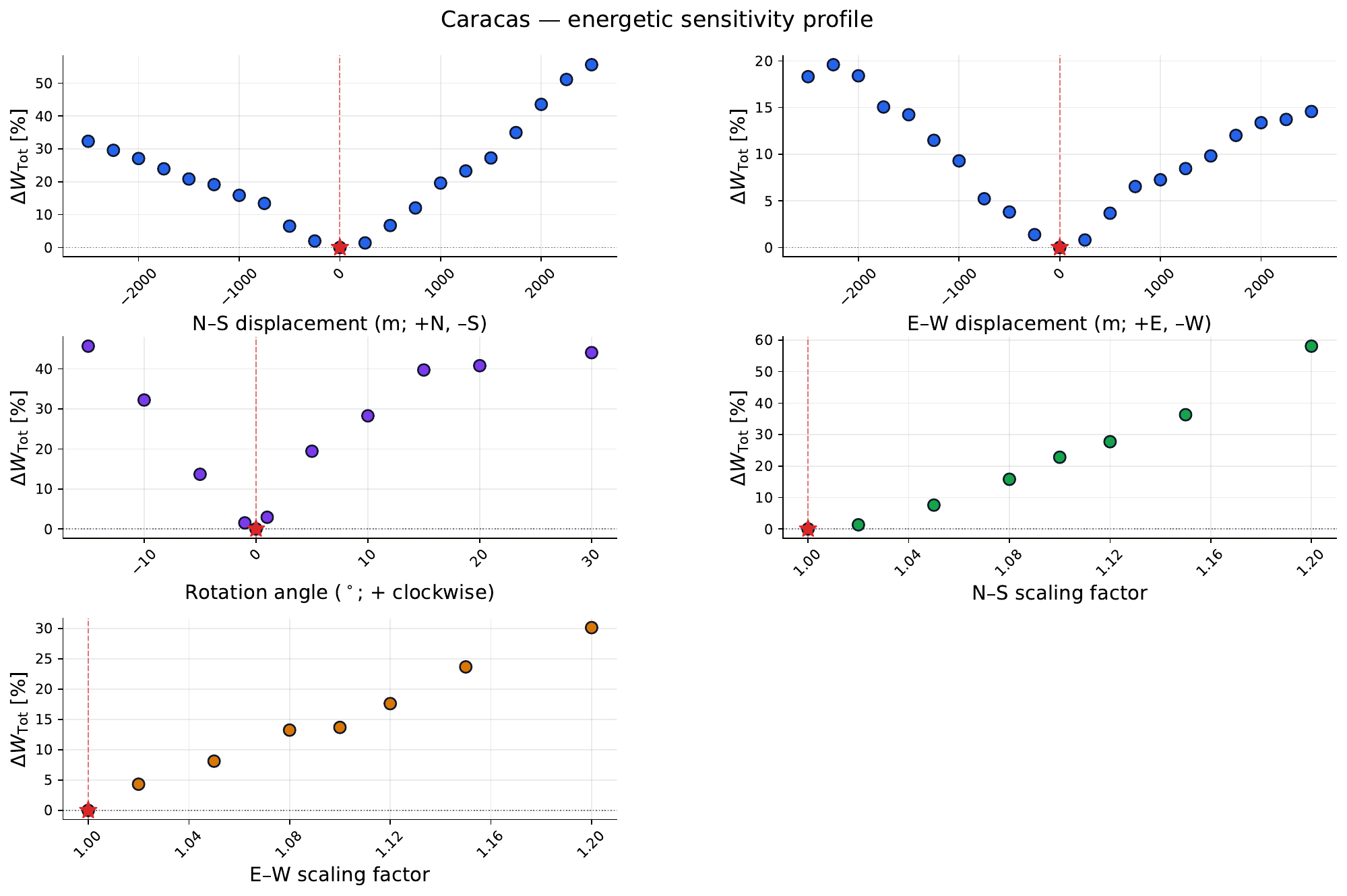}
\caption{\textbf{Caracas — energetic sensitivity profile.}}
\end{figure*}

\begin{figure*}[ht]
\centering
\includegraphics[width=0.96\textwidth]{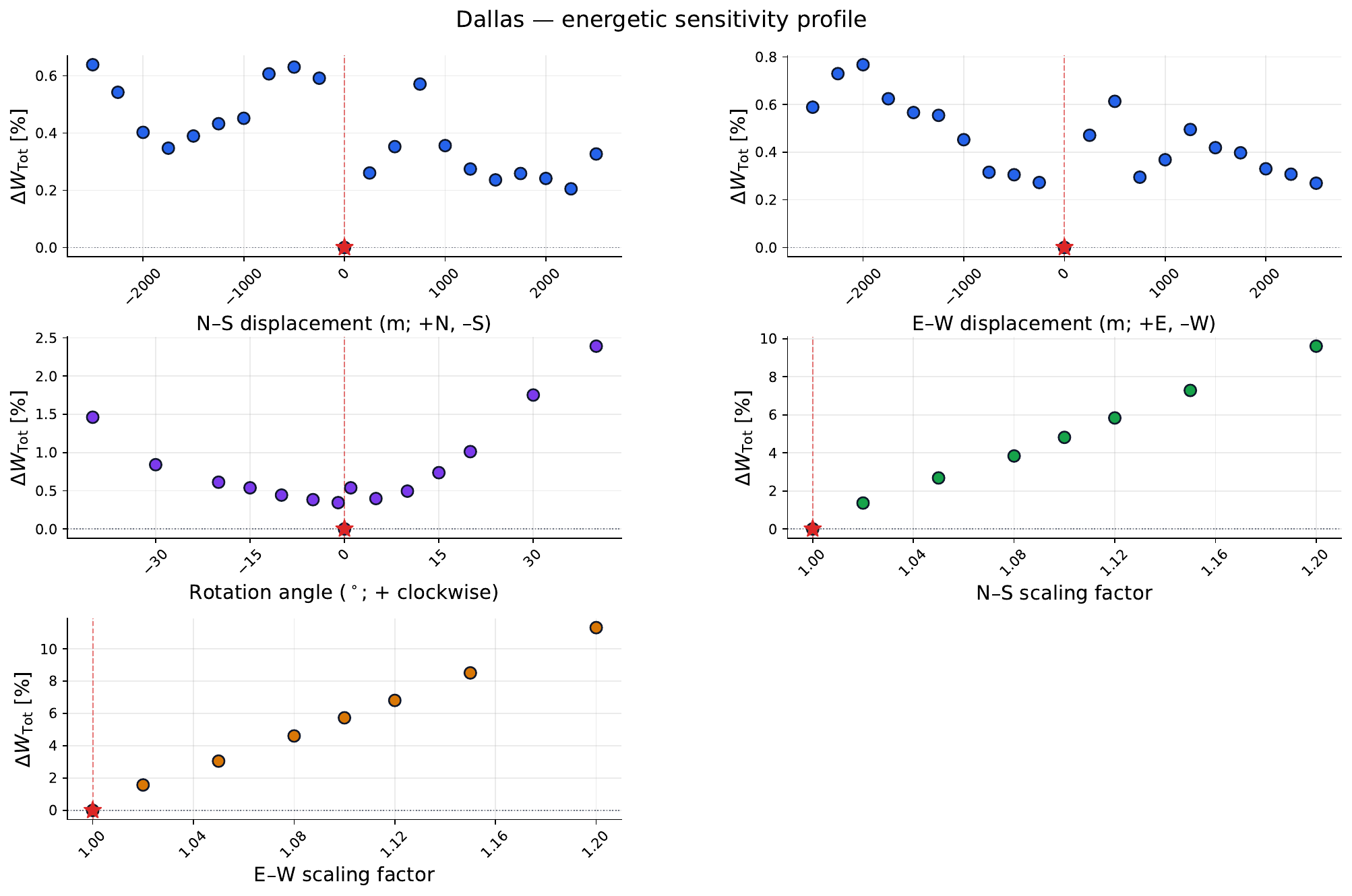}
\caption{\textbf{Dallas — energetic sensitivity profile.}}
\end{figure*}

\begin{figure*}[ht]
\centering
\includegraphics[width=0.96\textwidth]{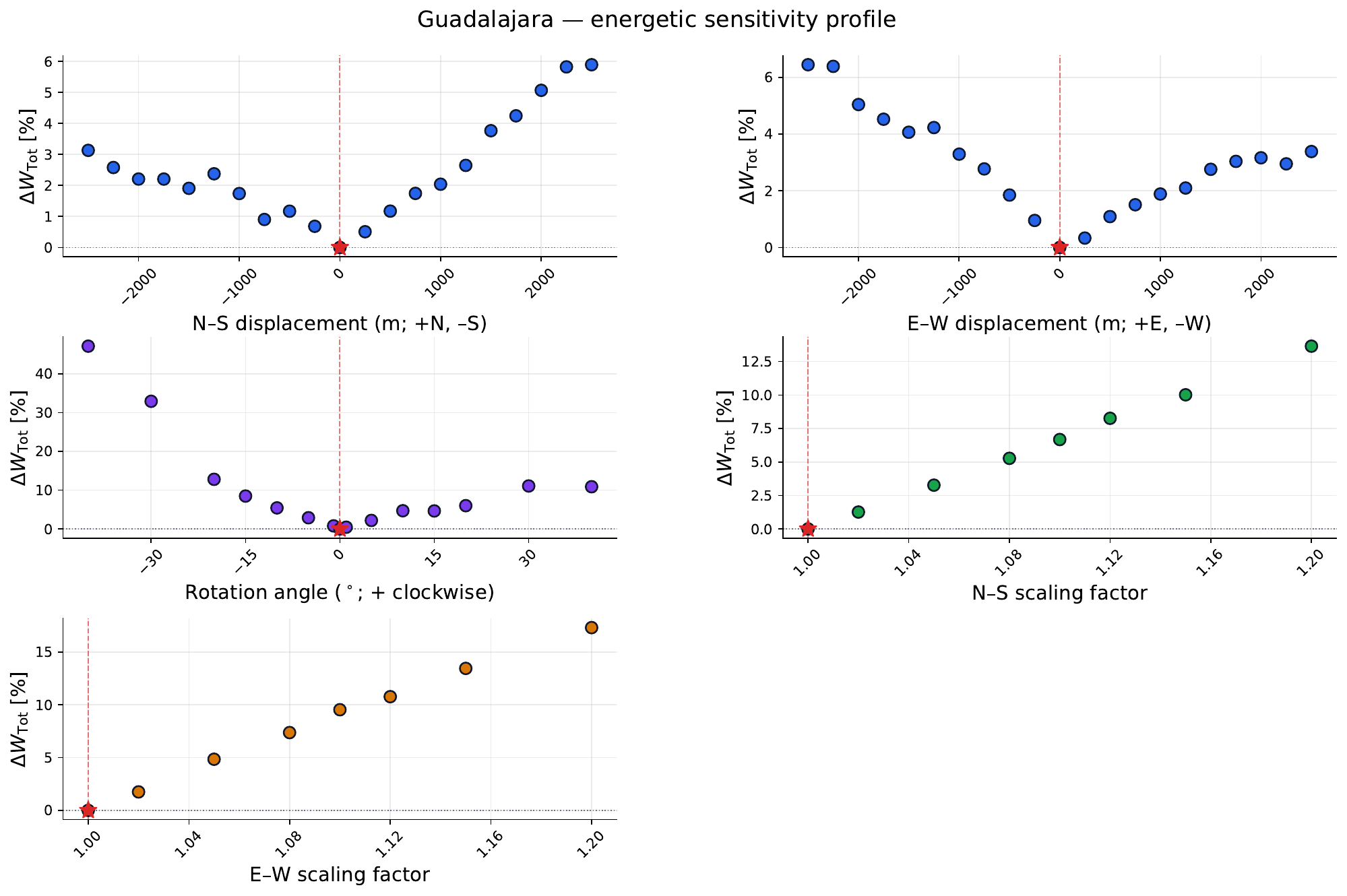}
\caption{\textbf{Guadalajara — energetic sensitivity profile.}}
\end{figure*}

\begin{figure*}[ht]
\centering
\includegraphics[width=0.96\textwidth]{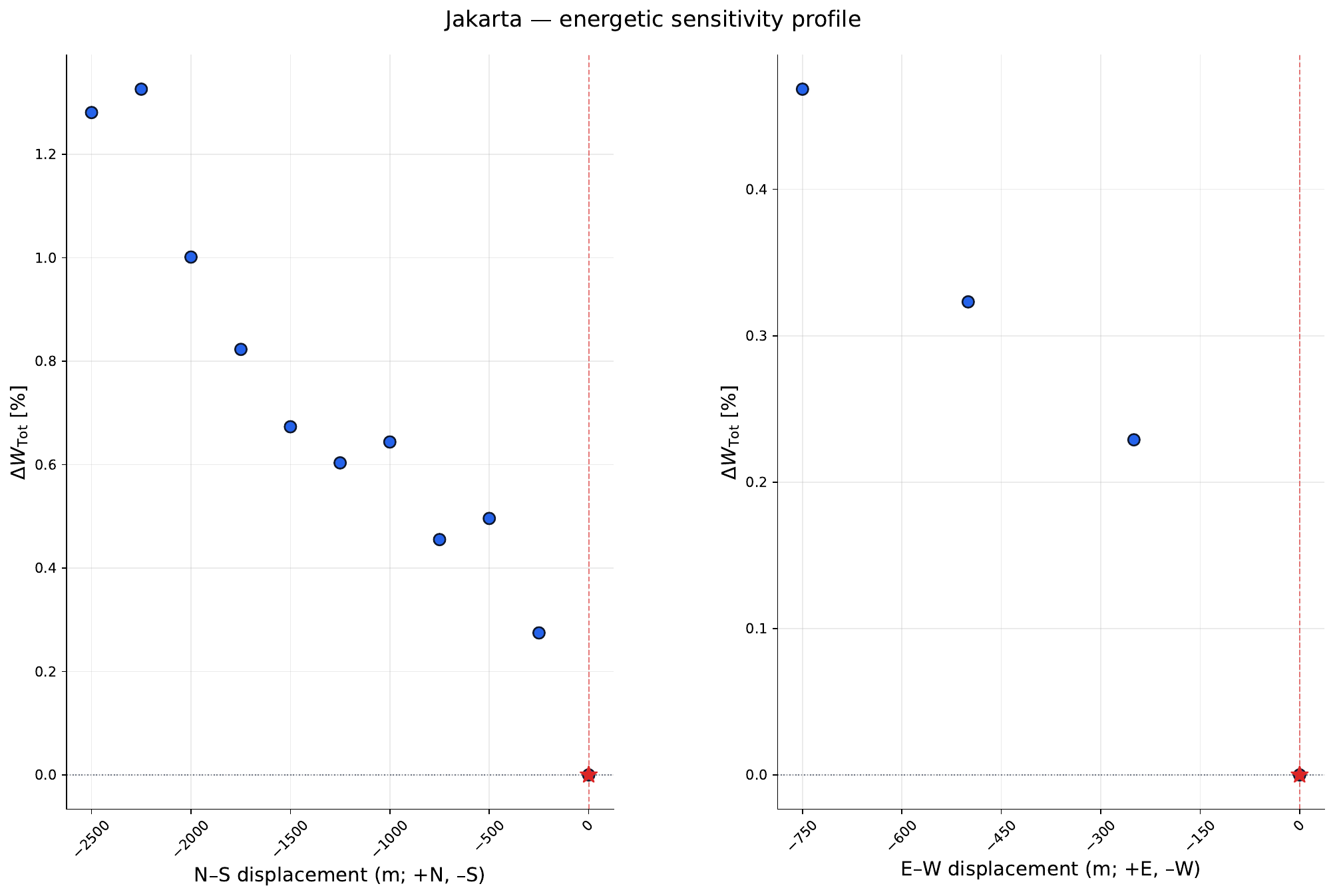}
\caption{\textbf{Jakarta — energetic sensitivity profile.}}
\end{figure*}

\begin{figure*}[ht]
\centering
\includegraphics[width=0.96\textwidth]{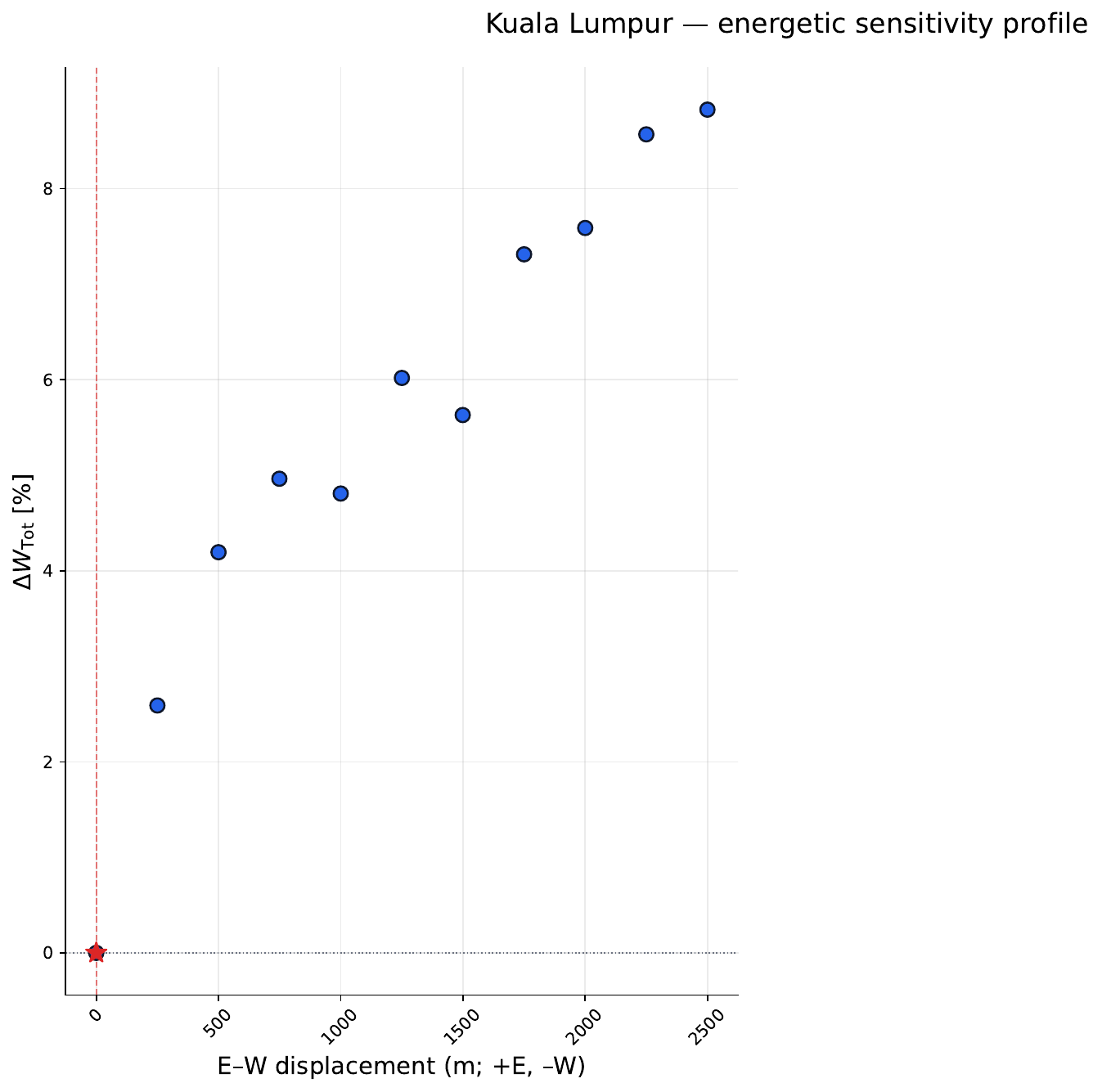}
\caption{\textbf{Kuala Lumpur — energetic sensitivity profile.}}
\end{figure*}

\begin{figure*}[ht]
\centering
\includegraphics[width=0.96\textwidth]{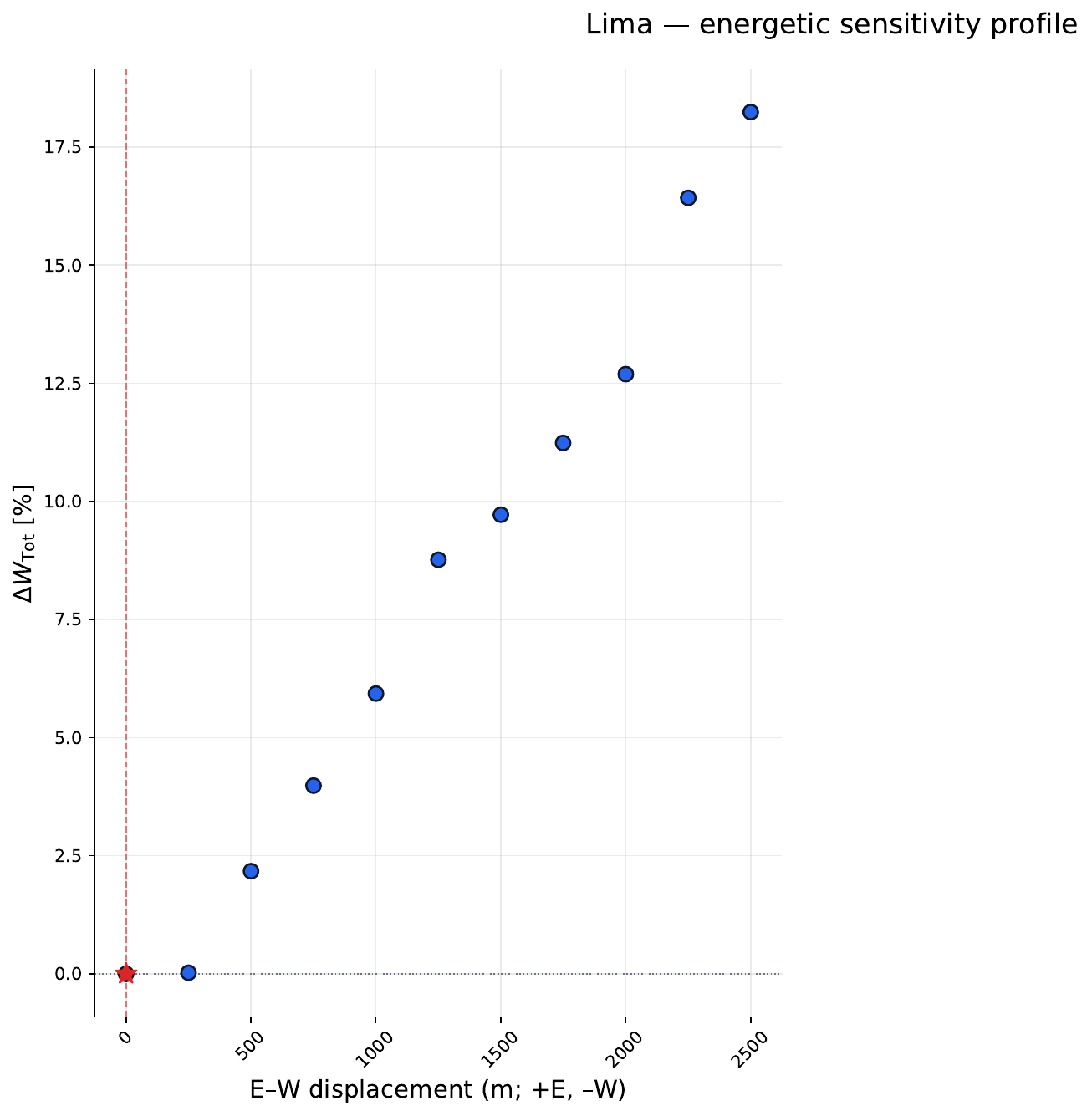}
\caption{\textbf{Lima — energetic sensitivity profile.}}
\end{figure*}

\begin{figure*}[ht]
\centering
\includegraphics[width=0.96\textwidth]{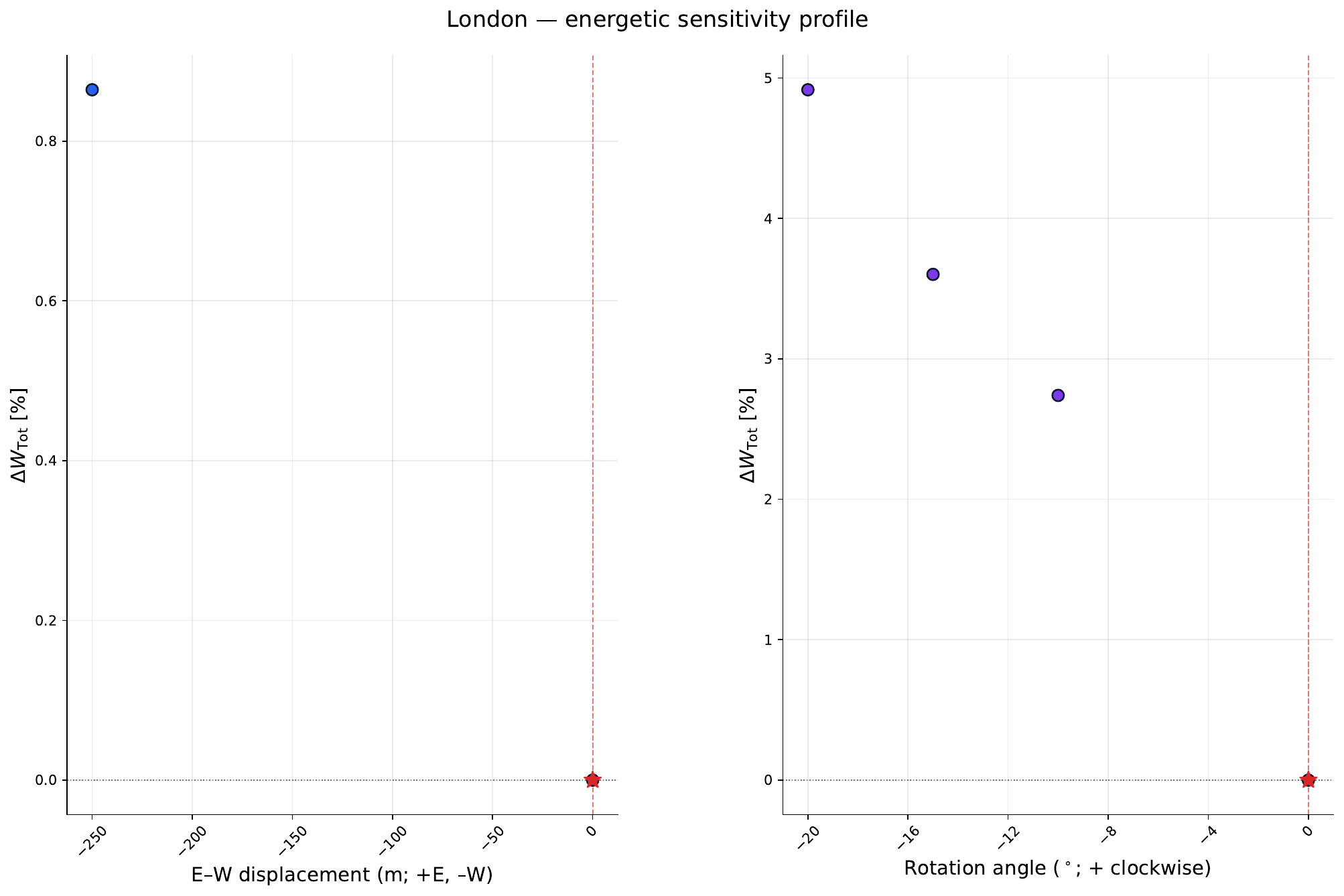}
\caption{\textbf{London — energetic sensitivity profile.}}
\end{figure*}

\begin{figure*}[ht]
\centering
\includegraphics[width=0.96\textwidth]{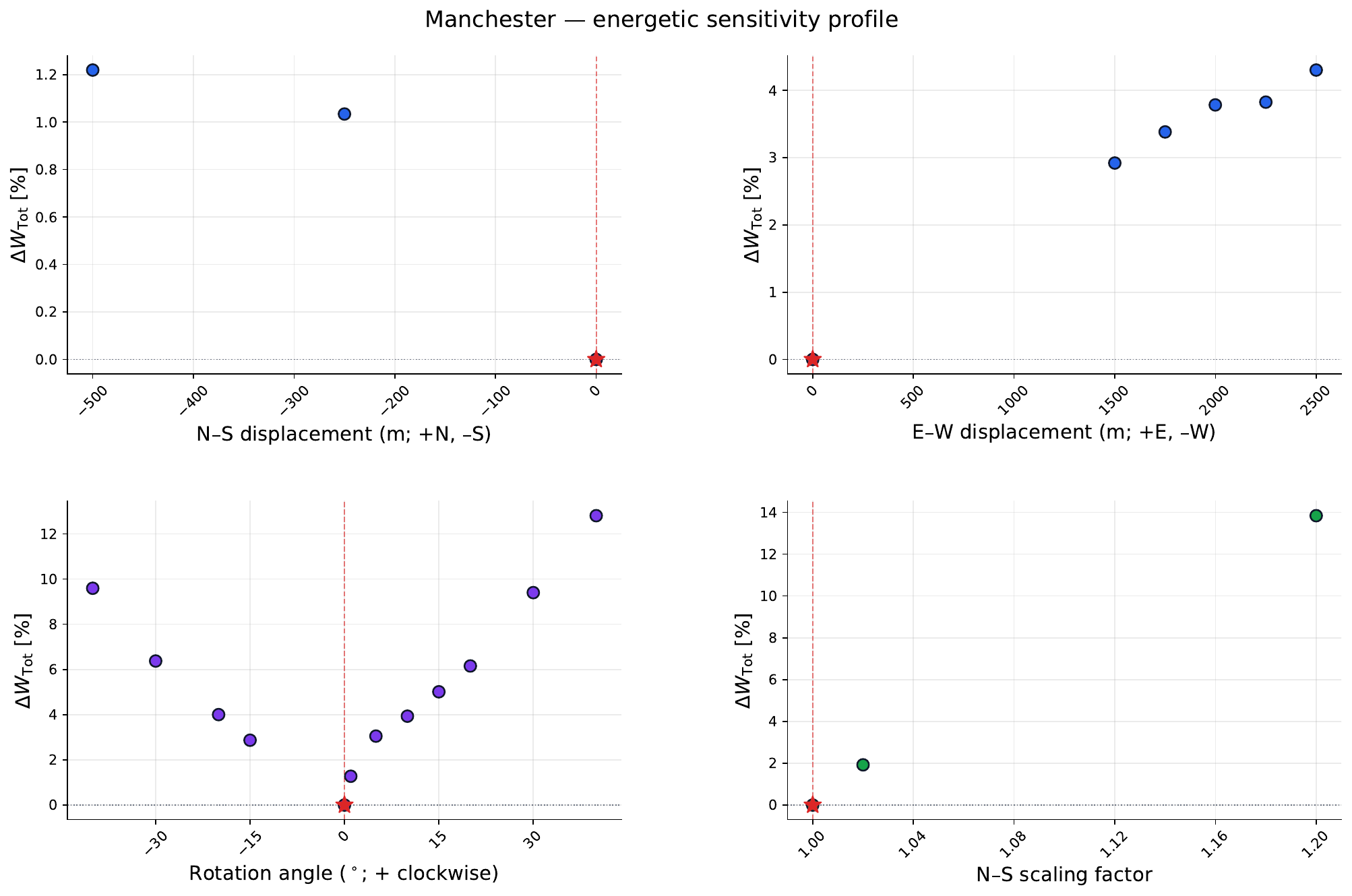}
\caption{\textbf{Manchester — energetic sensitivity profile.}}
\end{figure*}

\begin{figure*}[ht]
\centering
\includegraphics[width=0.96\textwidth]{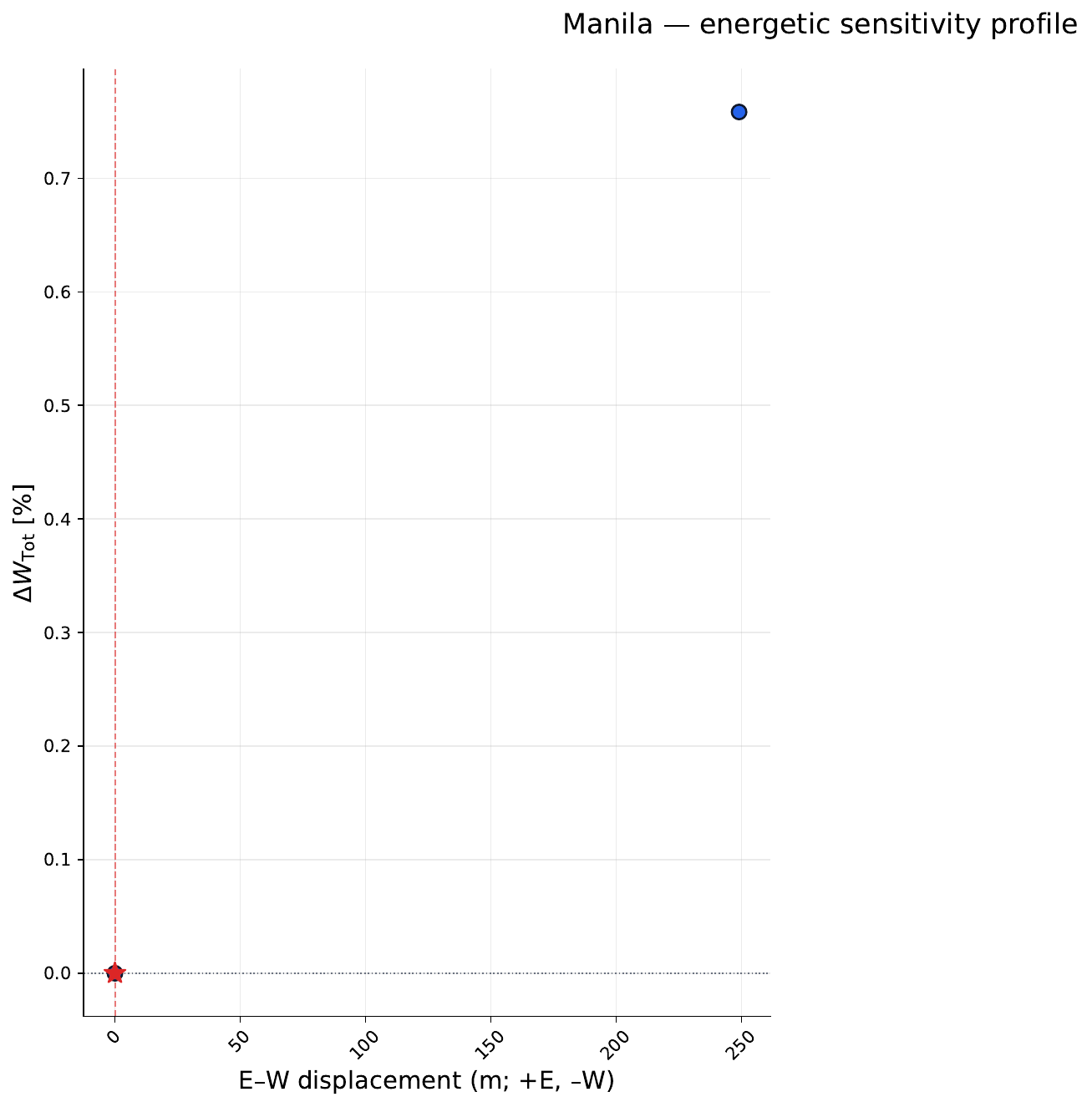}
\caption{\textbf{Manila — energetic sensitivity profile.}}
\end{figure*}

\begin{figure*}[ht]
\centering
\includegraphics[width=0.96\textwidth]{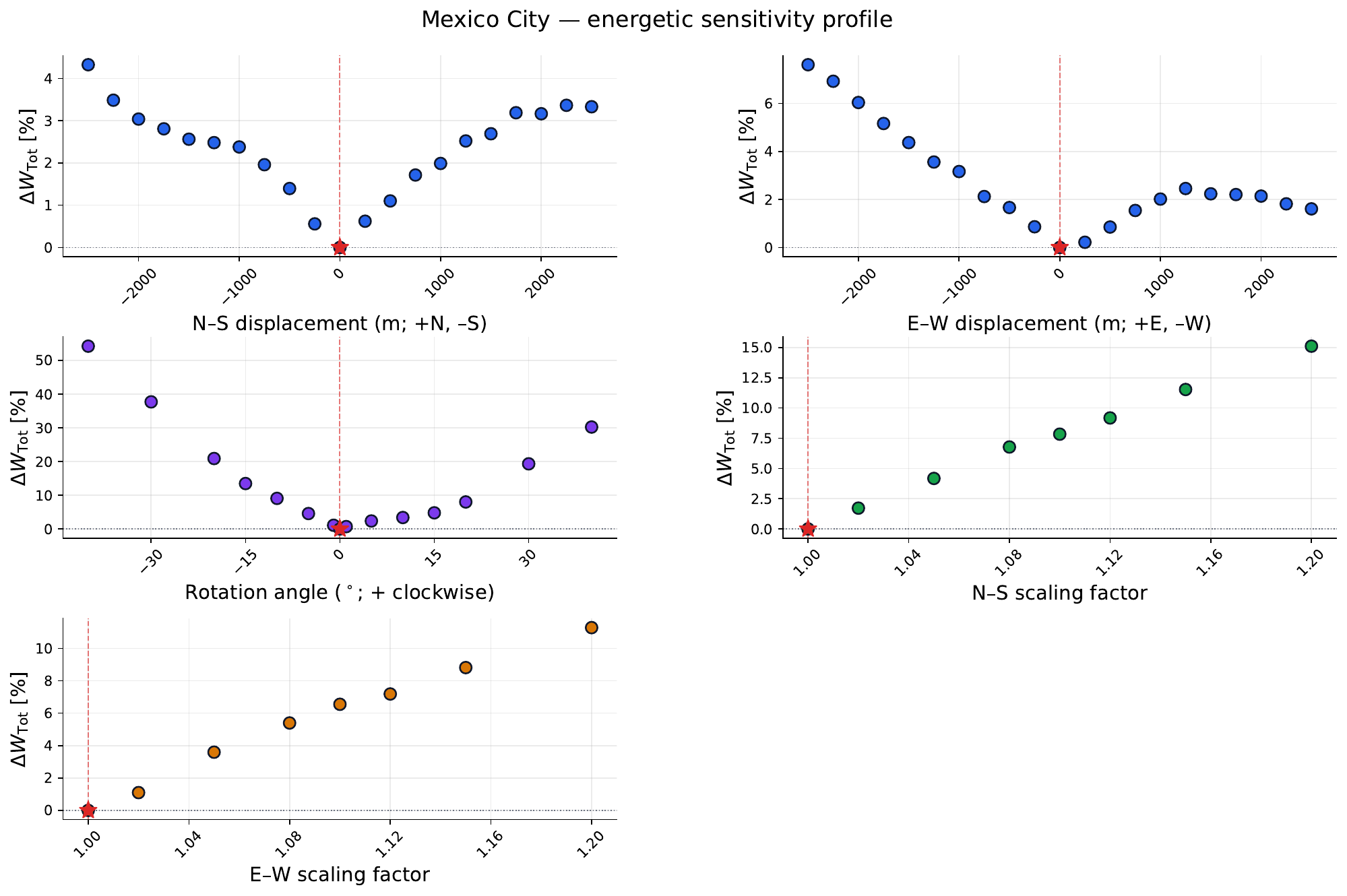}
\caption{\textbf{Mexico City — energetic sensitivity profile.}}
\end{figure*}

\begin{figure*}[ht]
\centering
\includegraphics[width=0.96\textwidth]{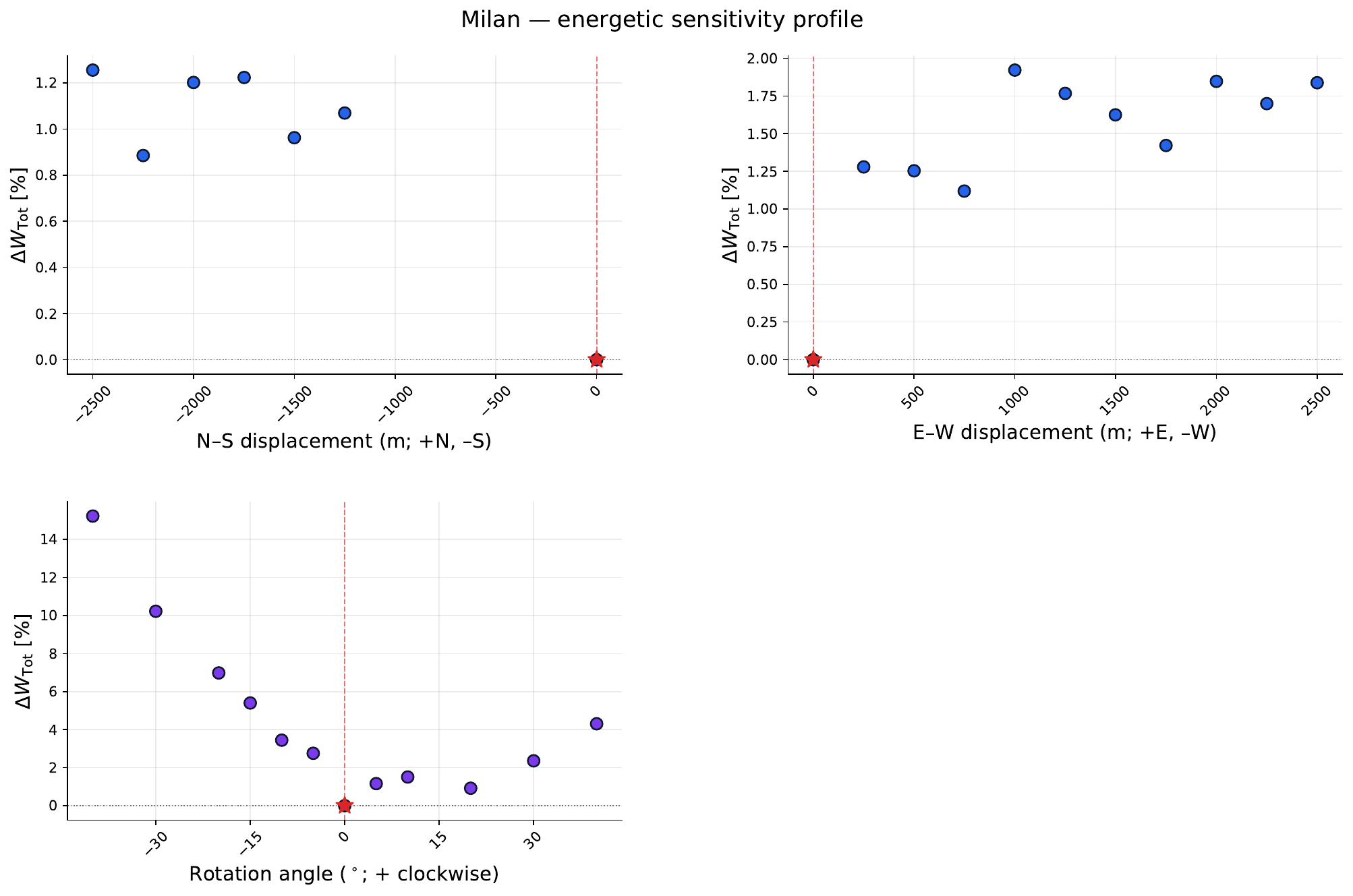}
\caption{\textbf{Milan — energetic sensitivity profile.}}
\end{figure*}

\begin{figure*}[ht]
\centering
\includegraphics[width=0.96\textwidth]{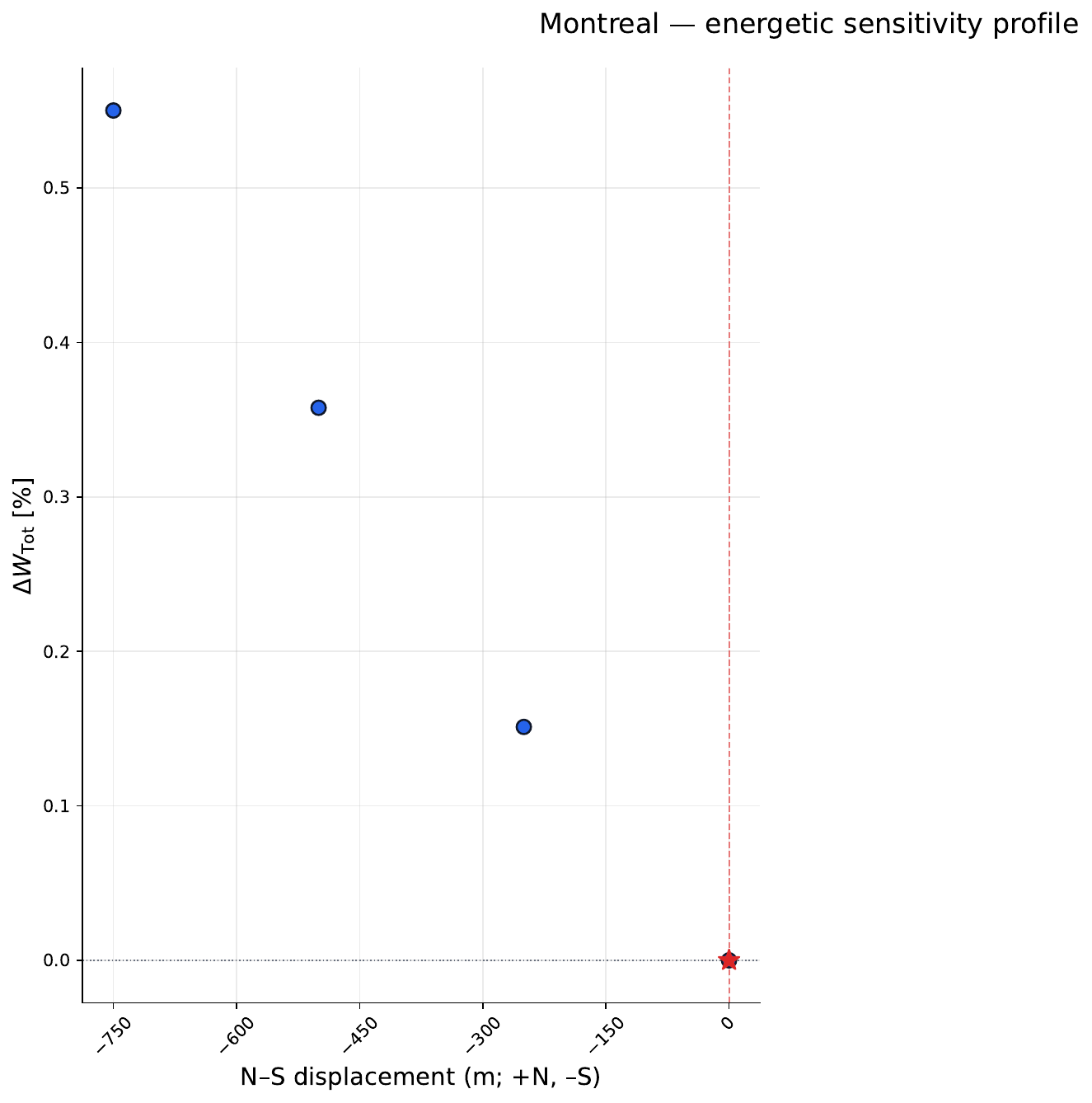}
\caption{\textbf{Montreal — energetic sensitivity profile.}}
\end{figure*}

\begin{figure*}[ht]
\centering
\includegraphics[width=0.96\textwidth]{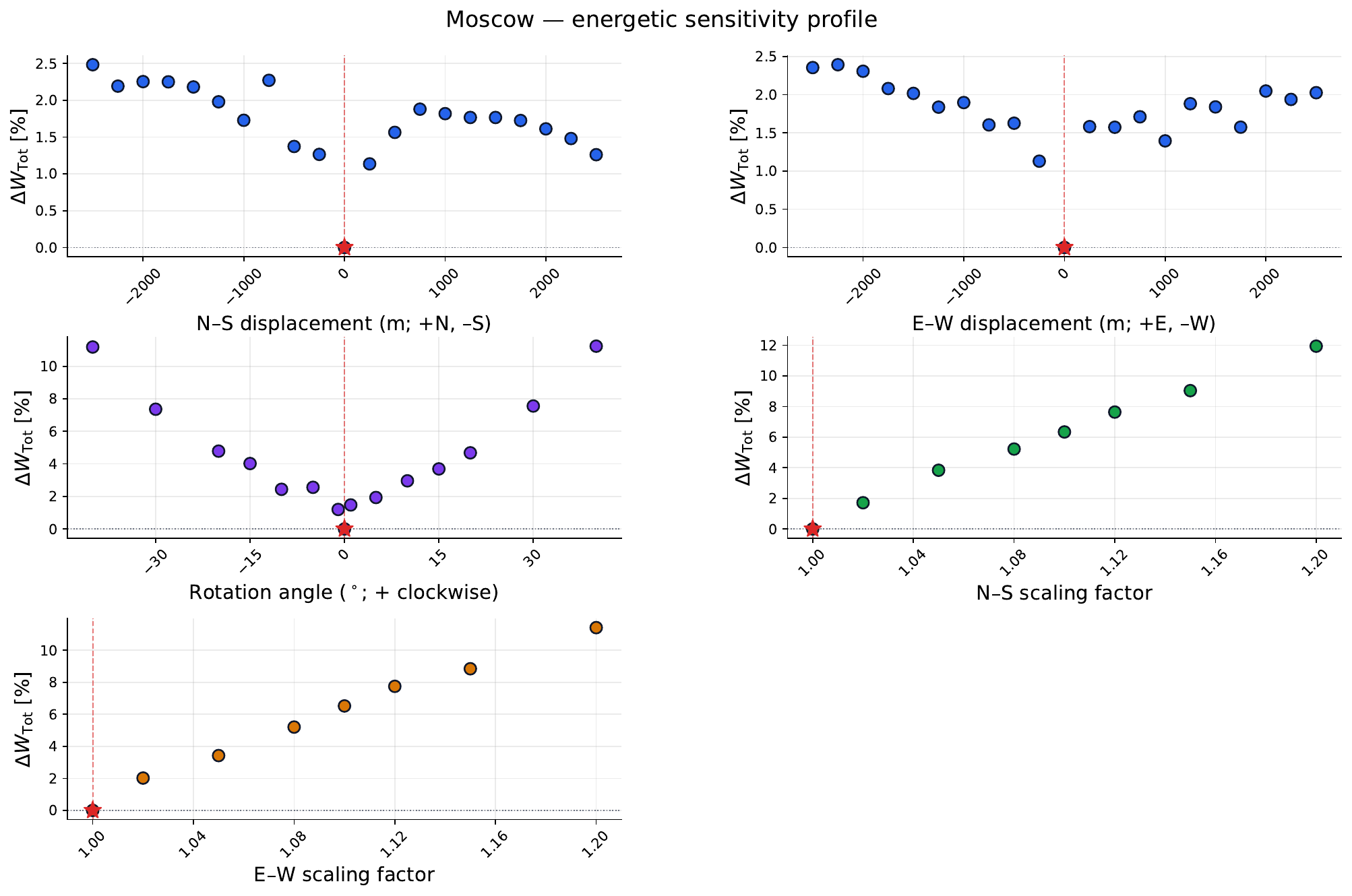}
\caption{\textbf{Moscow — energetic sensitivity profile.}}
\end{figure*}

\begin{figure*}[ht]
\centering
\includegraphics[width=0.96\textwidth]{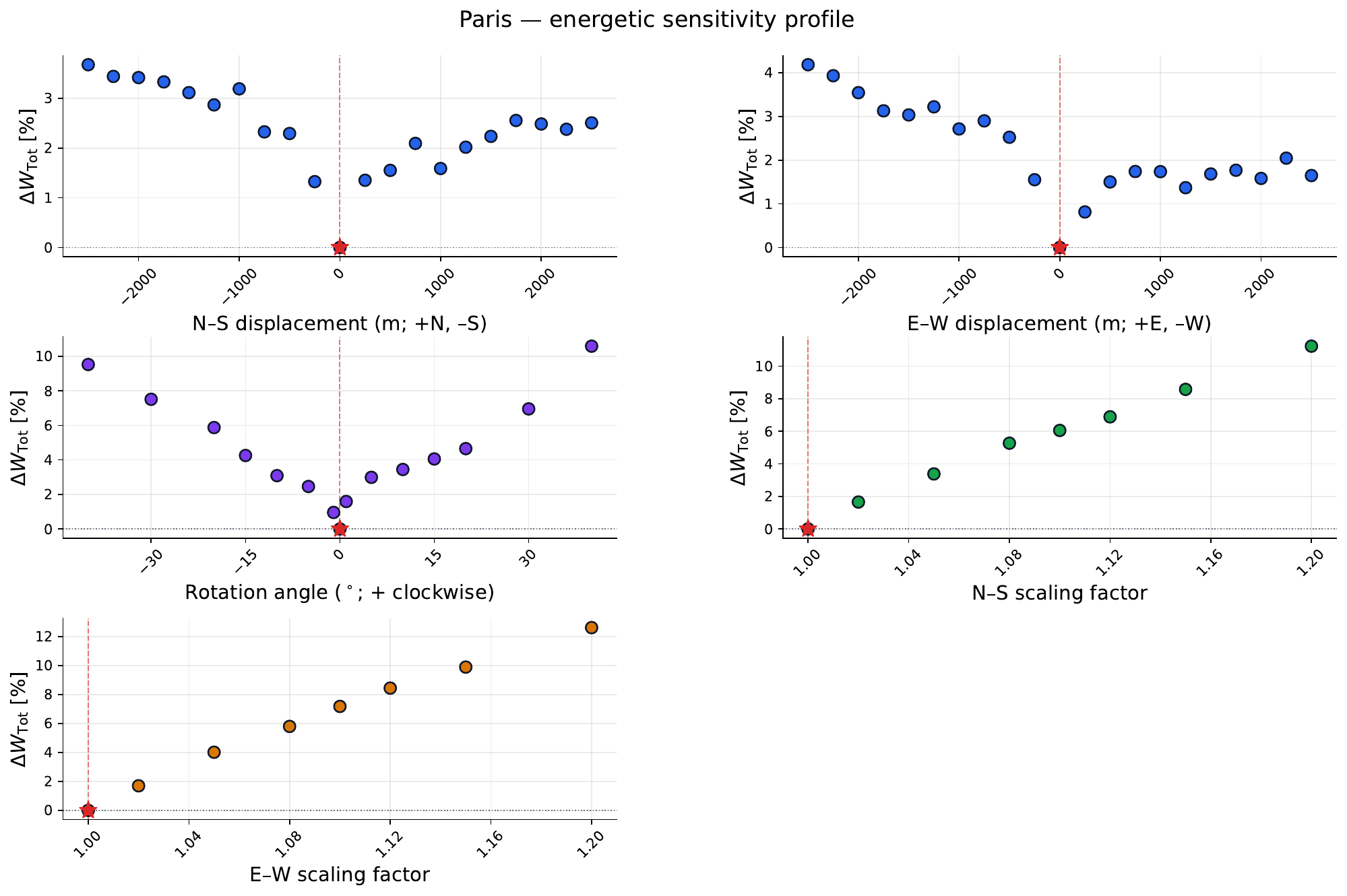}
\caption{\textbf{Paris — energetic sensitivity profile.}}
\end{figure*}

\begin{figure*}[ht]
\centering
\includegraphics[width=0.96\textwidth]{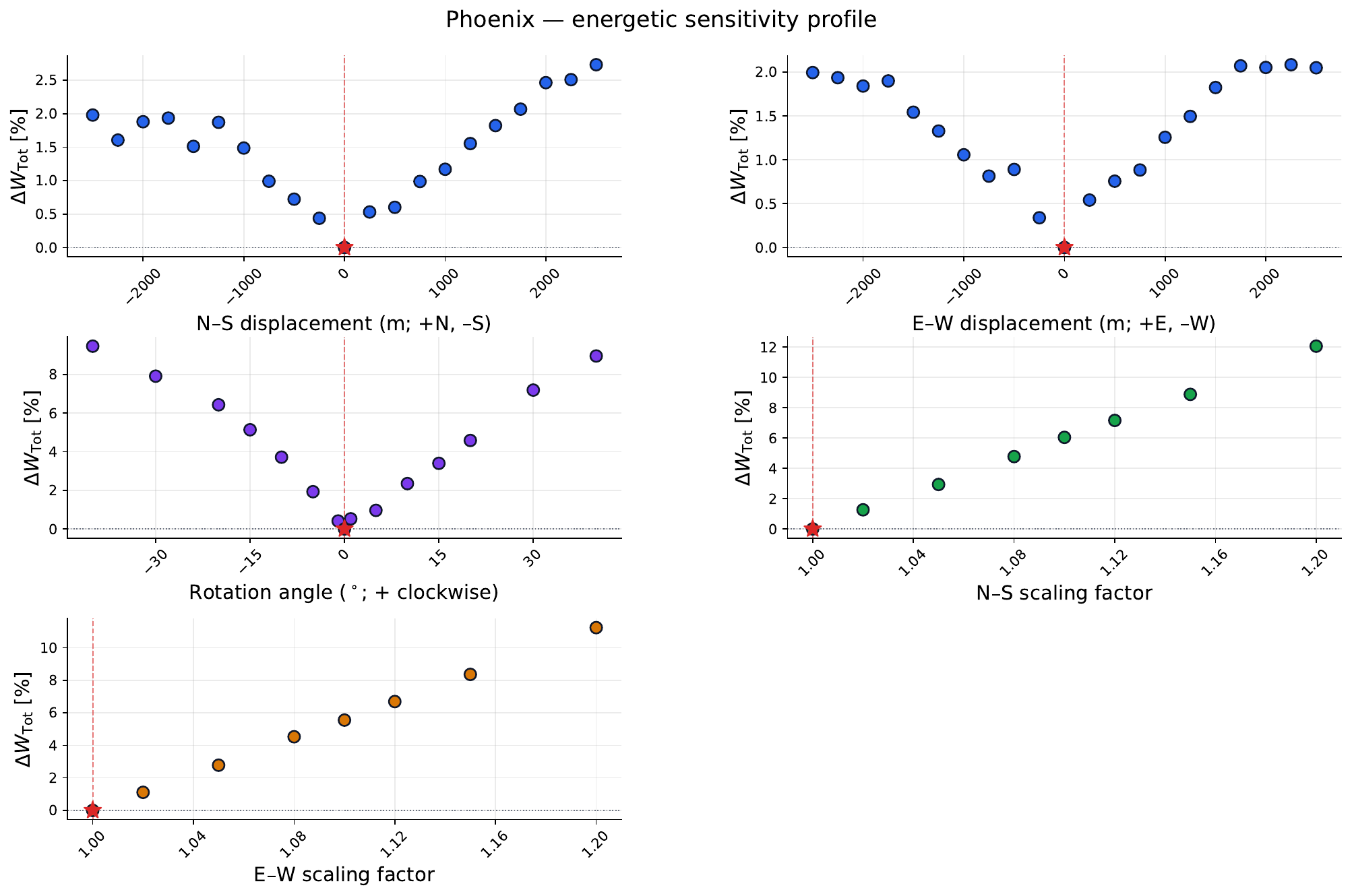}
\caption{\textbf{Phoenix — energetic sensitivity profile.}}
\end{figure*}

\begin{figure*}[ht]
\centering
\includegraphics[width=0.96\textwidth]{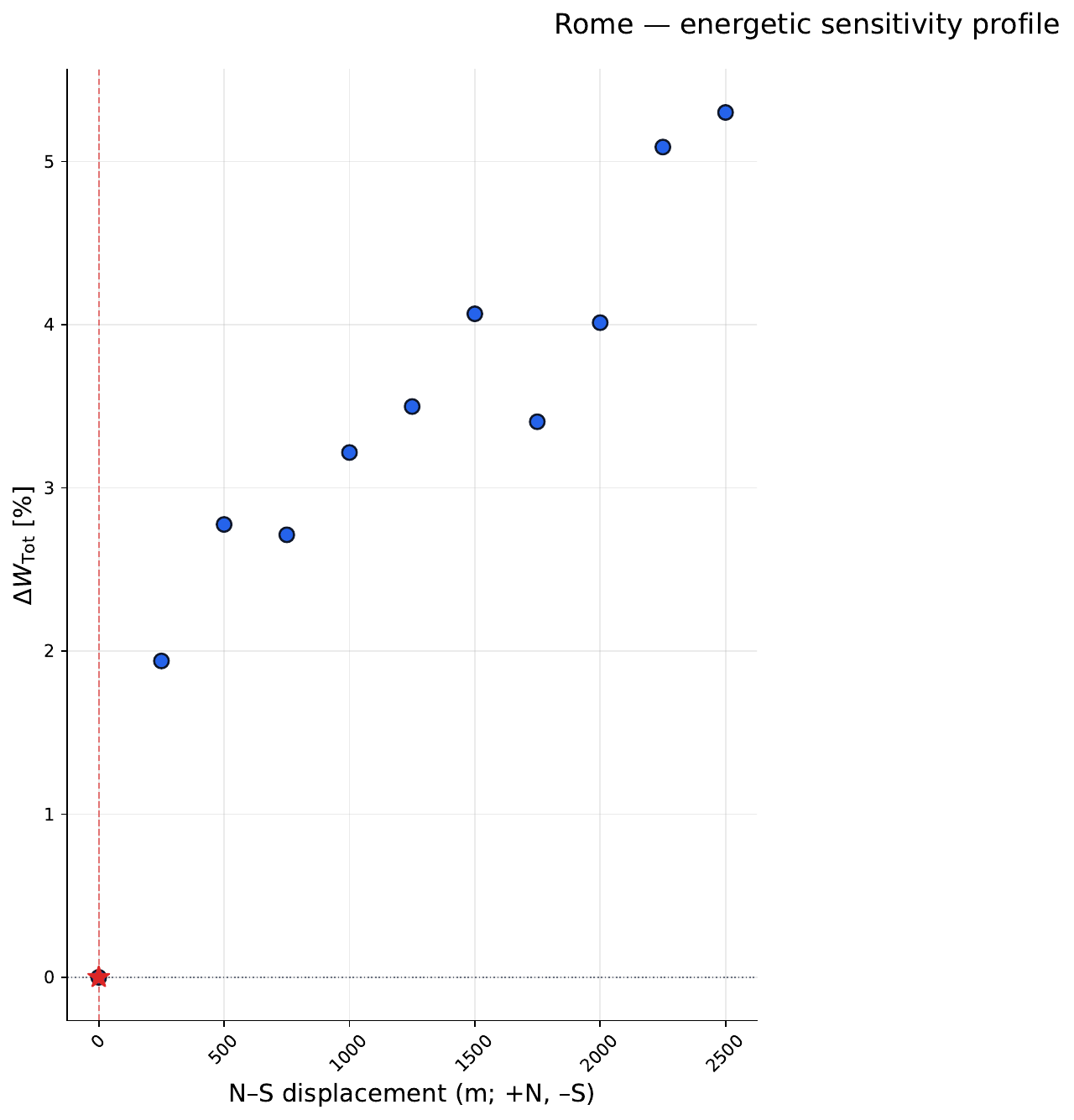}
\caption{\textbf{Rome — energetic sensitivity profile.}}
\end{figure*}

\begin{figure*}[ht]
\centering
\includegraphics[width=0.96\textwidth]{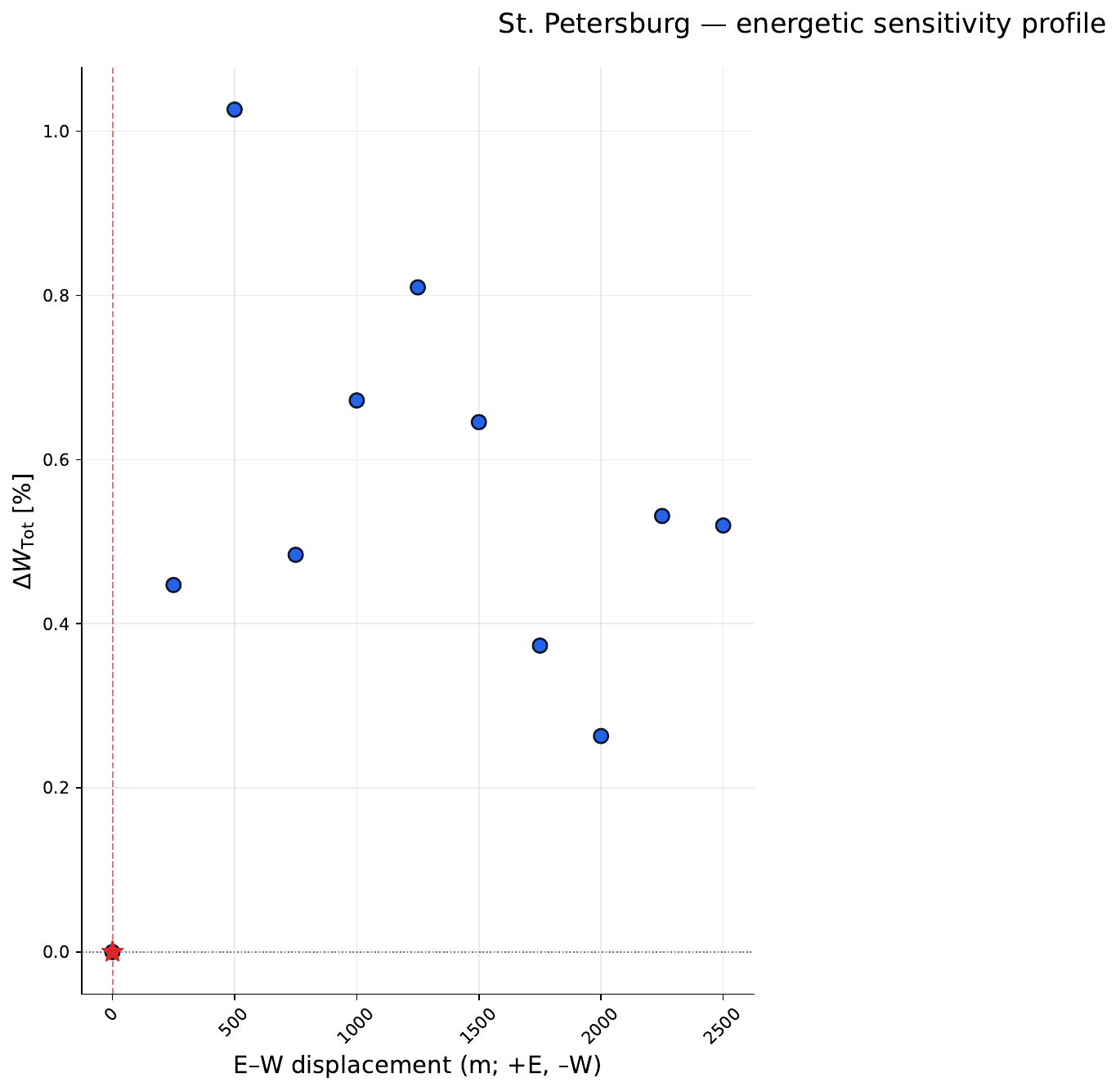}
\caption{\textbf{St. Petersburg — energetic sensitivity profile.}}
\end{figure*}

\begin{figure*}[ht]
\centering
\includegraphics[width=0.96\textwidth]{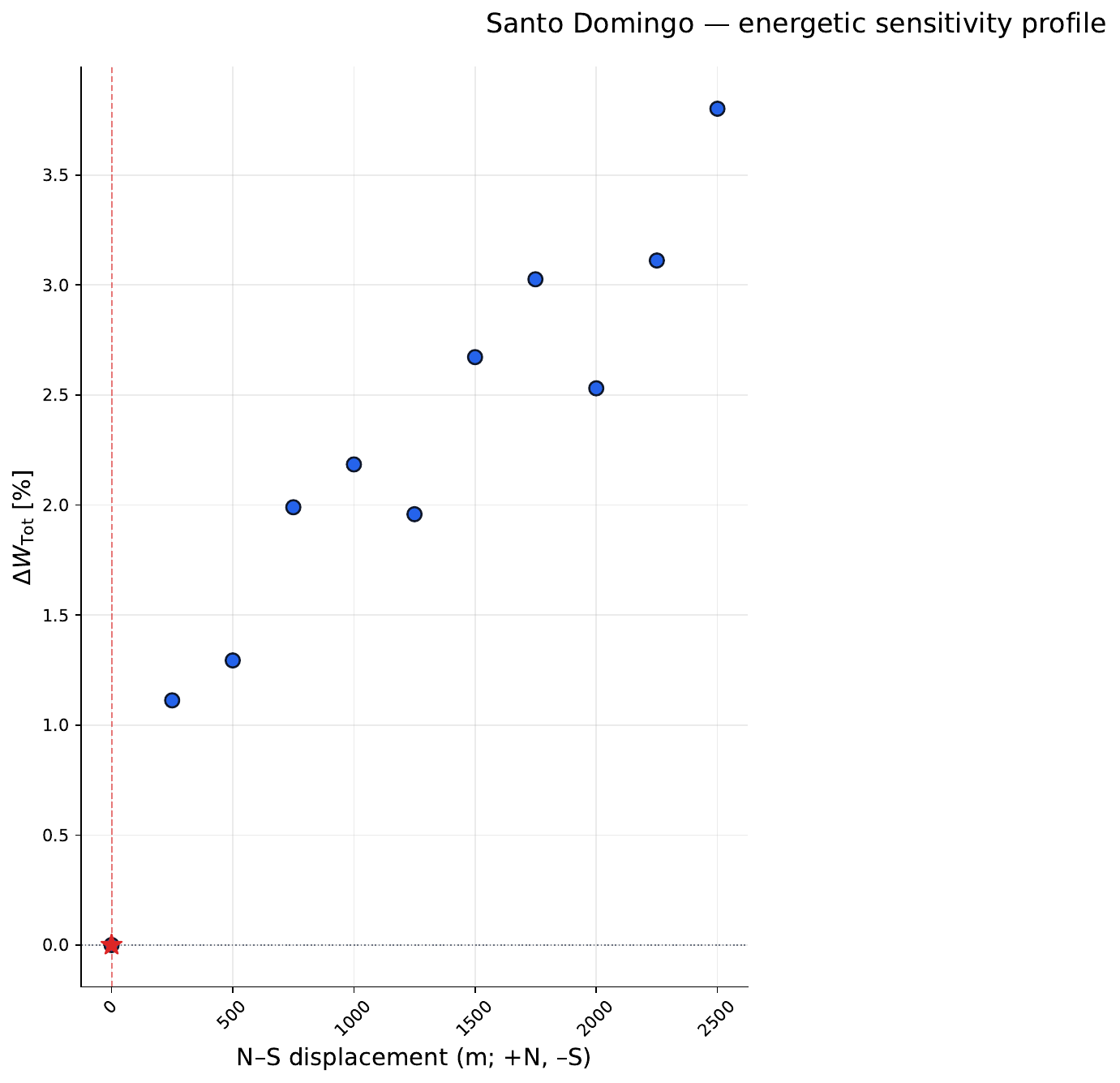}
\caption{\textbf{Santo Domingo — energetic sensitivity profile.}}
\end{figure*}

\begin{figure*}[ht]
\centering
\includegraphics[width=0.96\textwidth]{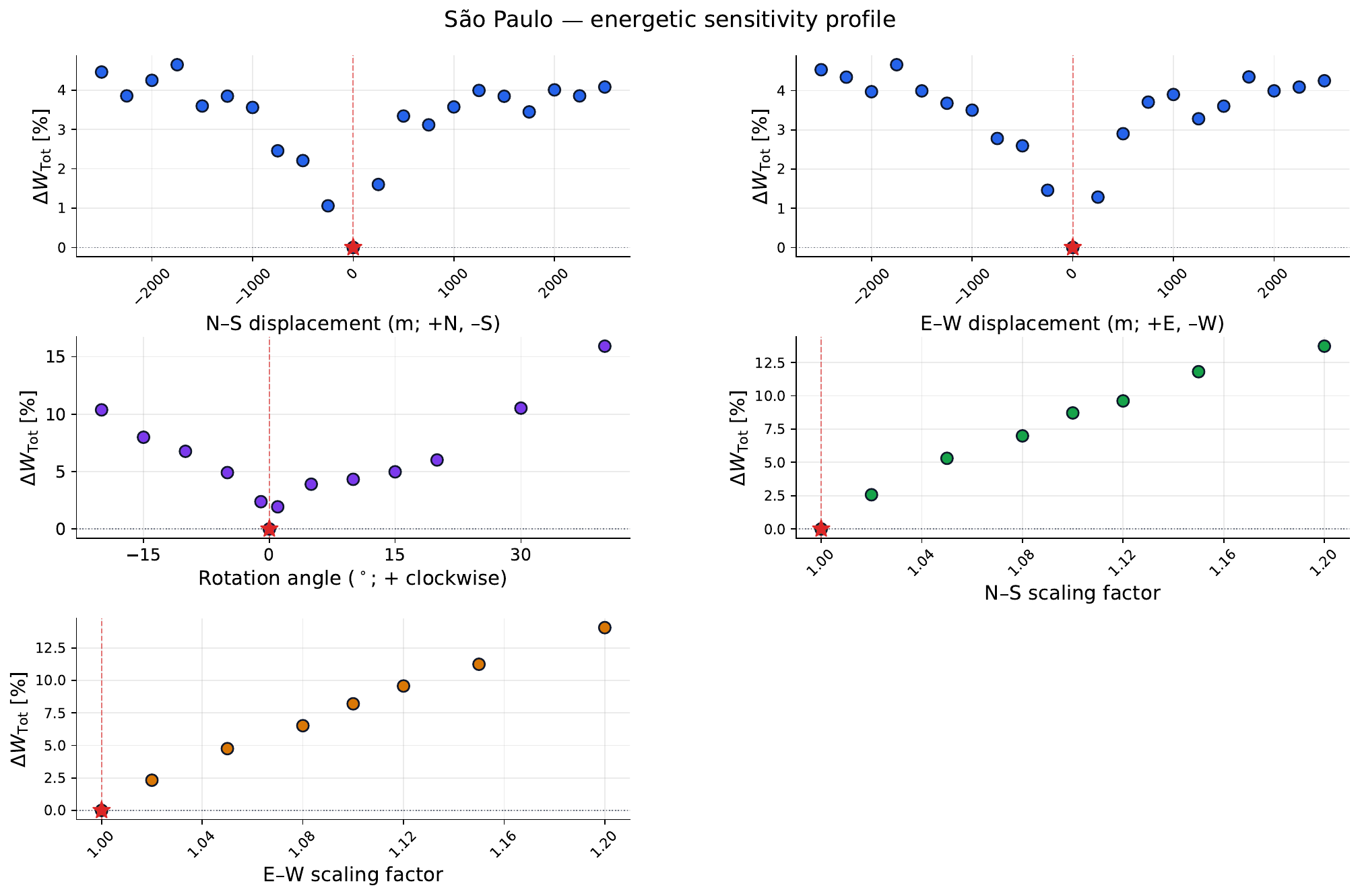}
\caption{\textbf{S\~ao Paulo — energetic sensitivity profile.}}
\end{figure*}

\begin{figure*}[ht]
\centering
\includegraphics[width=0.96\textwidth]{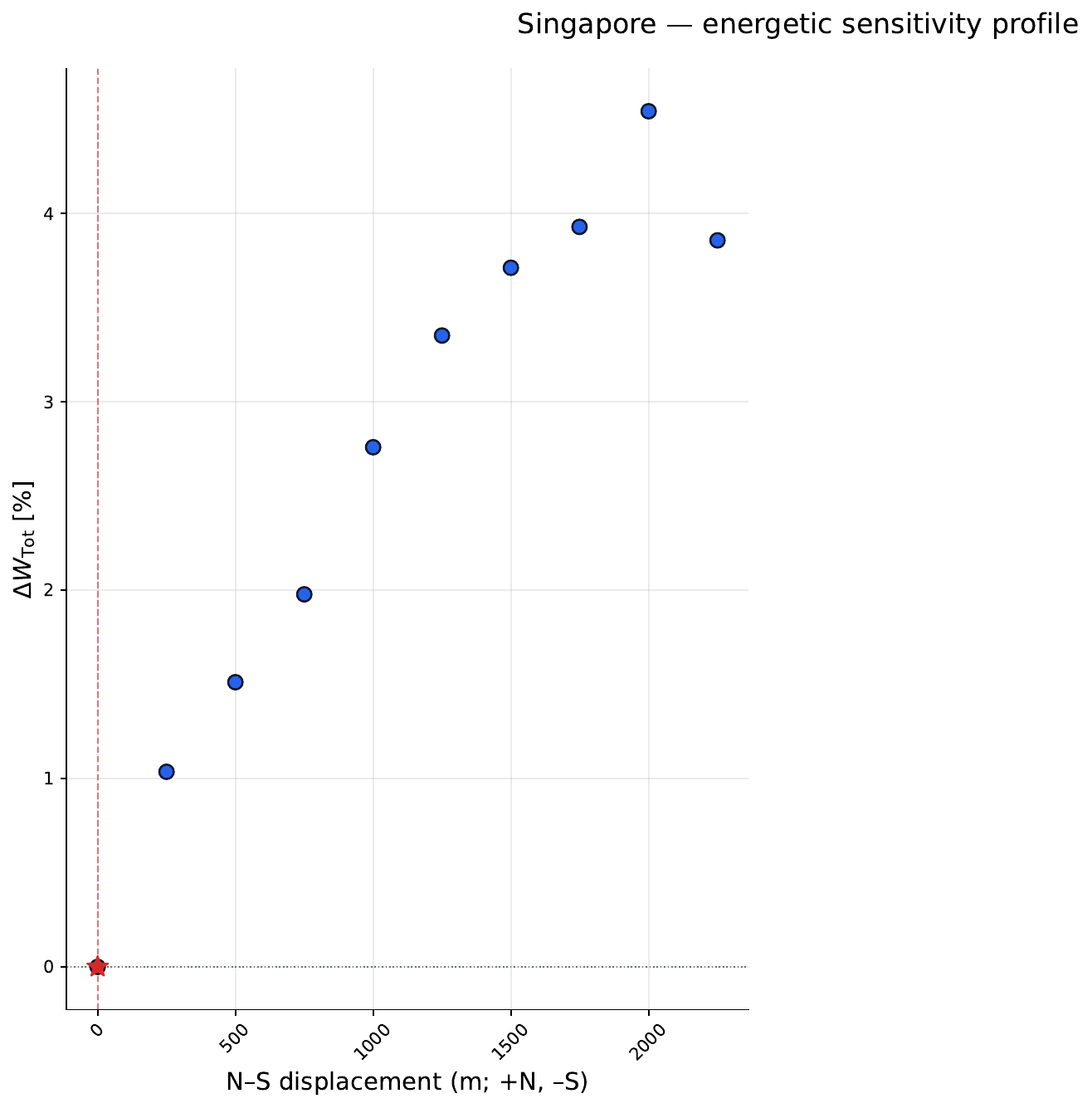}
\caption{\textbf{Singapore — energetic sensitivity profile.}}
\end{figure*}

\begin{figure*}[ht]
\centering
\includegraphics[width=0.96\textwidth]{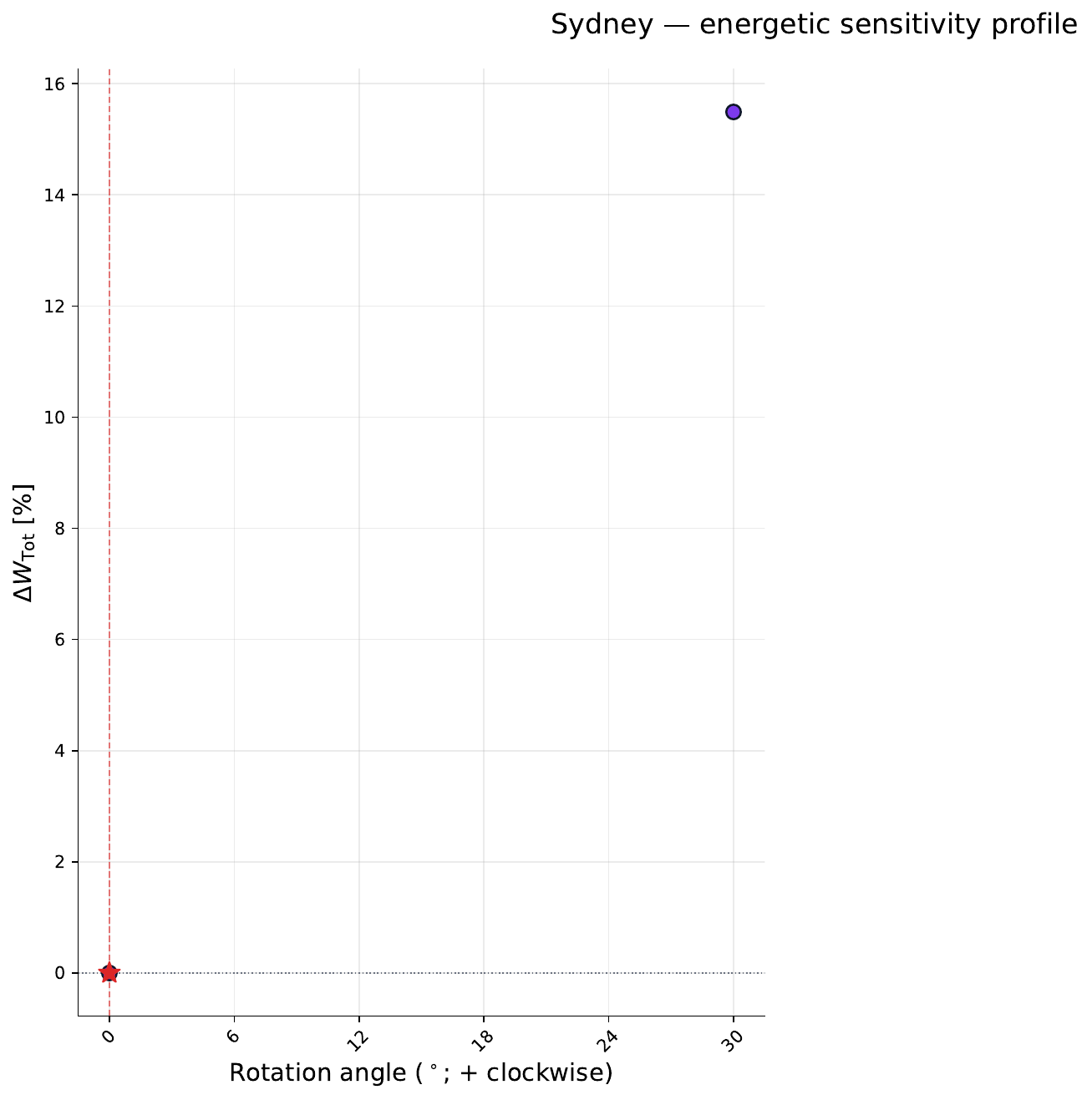}
\caption{\textbf{Sydney — energetic sensitivity profile.}}
\end{figure*}

\section*{Supplementary Note 4: Cities where a displaced configuration achieves lower energy}
\label{sec2}

\begin{figure*}[p]
\centering
\includegraphics[width=\textwidth]{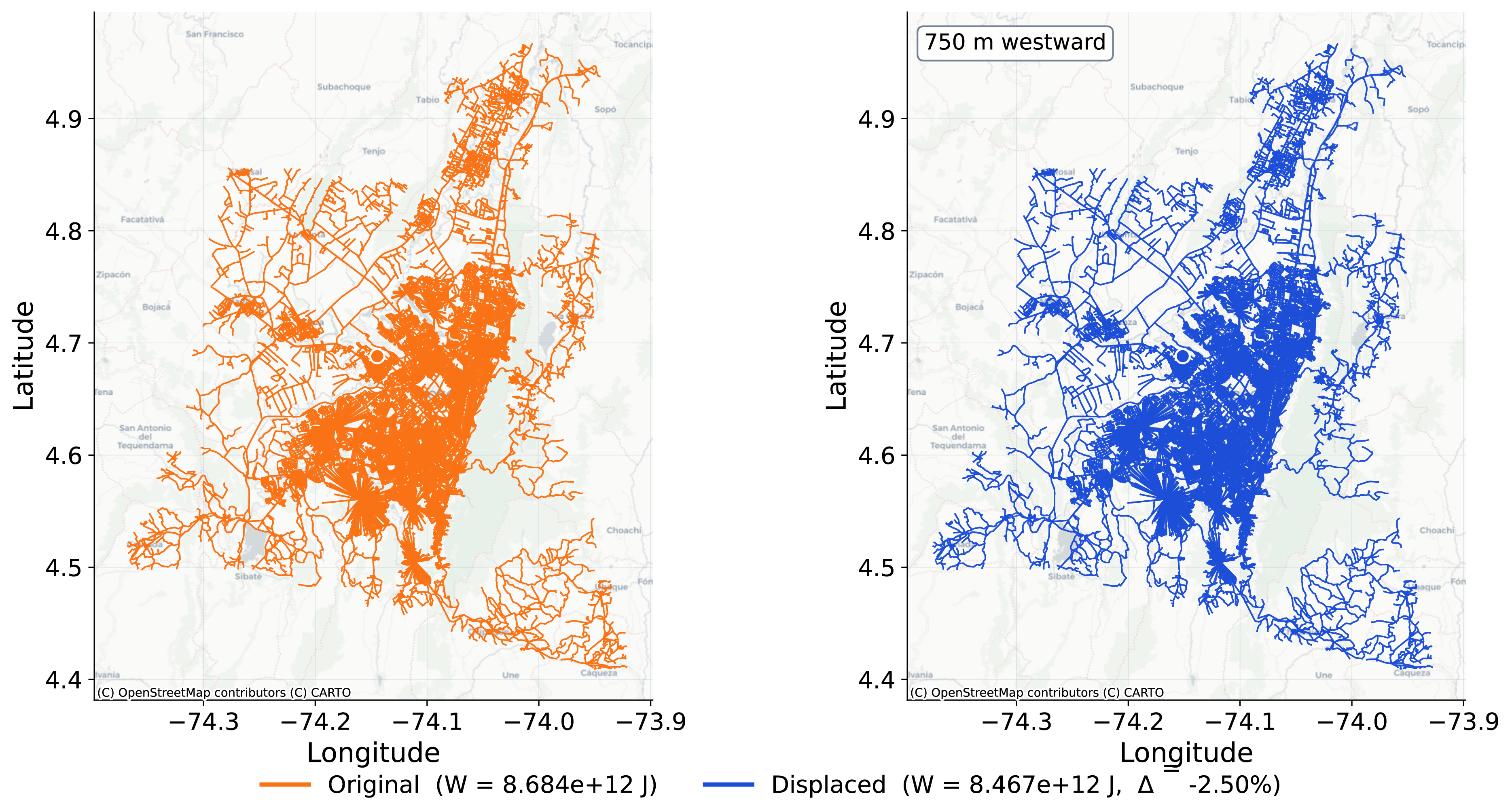}

\vspace{0.5cm}

\includegraphics[width=\textwidth]{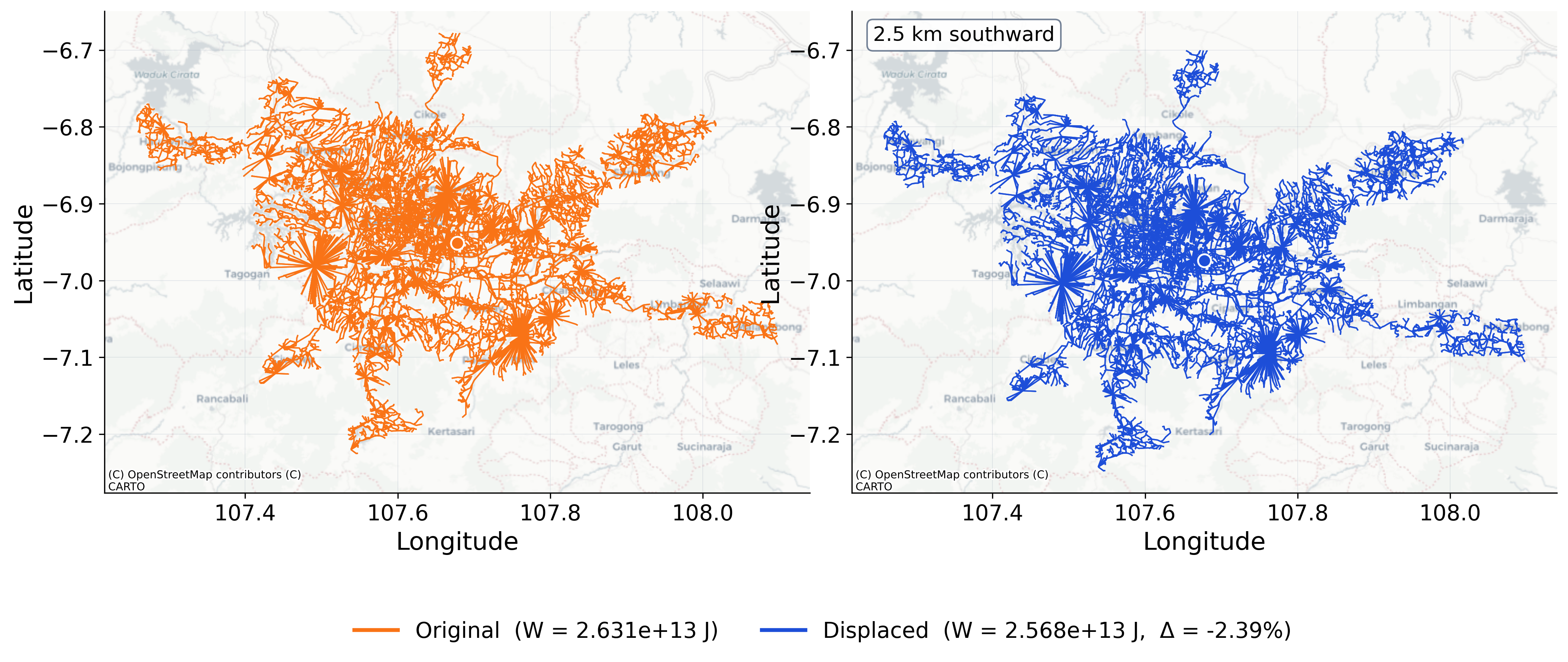}
\caption{\label{fig:supp:bogota_bandung}\textbf{Lower-energy displaced configurations — Bogot\'a (top) and Bandung (bottom).} Each panel shows the empirical road network (left) and the lowest-energy displaced configuration (right) over the same geographic extent. Orange: empirical embedding. Blue: displaced embedding.}
\end{figure*}

\begin{figure*}[p]
\centering
\includegraphics[width=\textwidth]{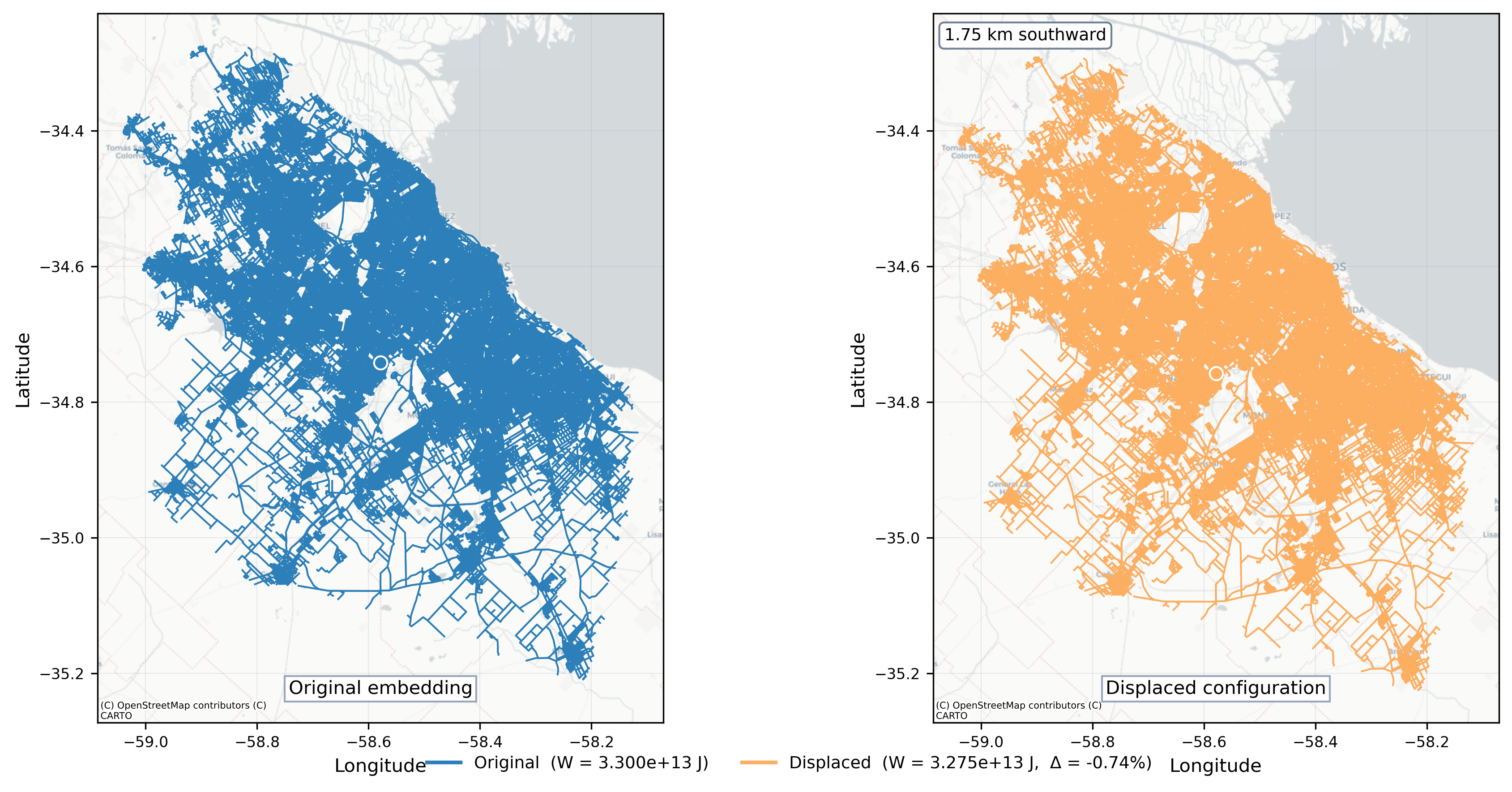}

\vspace{0.5cm}

\includegraphics[width=\textwidth]{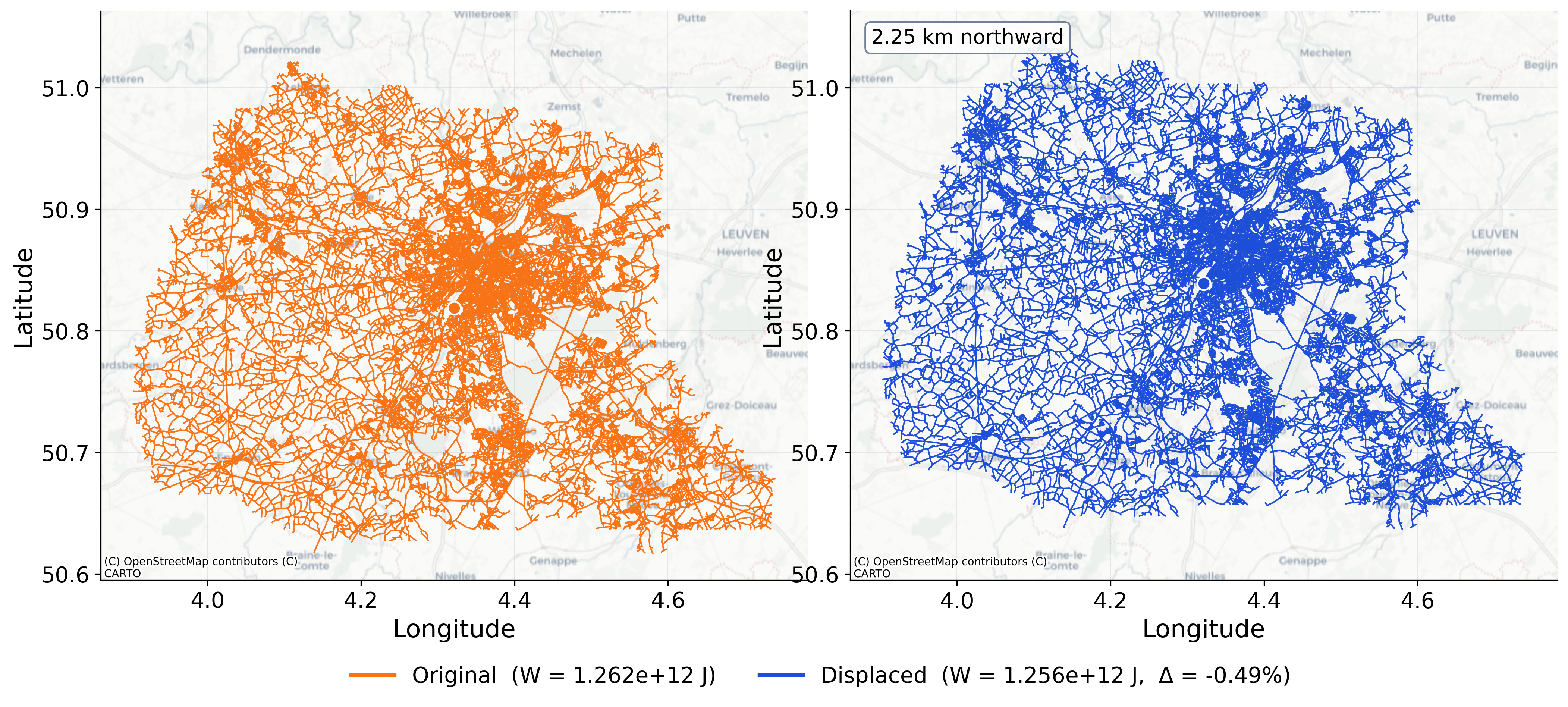}
\caption{\label{fig:supp:buenosaires_bruxelles}\textbf{Lower-energy displaced configurations — Buenos Aires (top) and Brussels (bottom).} Panels as in Supplementary Fig.~\ref{fig:supp:bogota_bandung}, except that in the Buenos Aires panels the empirical embedding is shown in blue and the displaced embedding in orange.}
\end{figure*}

In 4 cities, at least one displaced configuration achieves lower $W_{\mathrm{Tot}}$ than the observed layout. In all cases the deviation is below $3\%$ and a geographic rationale can be identified. These cases are consistent with identifiable spatial constraints and represent only small deviations from the overall low-energy pattern. Supplementary Figs.~\ref{fig:supp:bogota_bandung} and \ref{fig:supp:buenosaires_bruxelles} show the empirical and best-displaced configurations for each city.

\paragraph{Bandung ($\Delta = -2.39\%$, Supplementary Fig.~\ref{fig:supp:bogota_bandung} bottom).} Bandung occupies the Cekungan Bandung, a volcanic intra-montane basin at approximately $700\, \mathrm{m}$ elevation, enclosed by a ring of Quaternary volcanic peaks. The northern sector of the basin is dominated by the Tangkuban Perahu stratovolcano ($\sim 2100 \, \mathrm{m}$) and the Burangrang massif, which present an abrupt terrain boundary that generates high altitudinal costs on routes crossing the northern FUA edge. The southern and southeastern basin boundary is comparatively more gradual. The best-displaced configuration is a southward translation of $2500\, \mathrm{m}$, which shifts the road network away from the steep northern volcanic rim and toward the more gently sloping southern basin floor. This translation lies at the edge of the tested displacement range, so larger southward shifts were not evaluated. Bandung’s historical center developed near the northern basin floor, following colonial-era infrastructure built along the main access corridor from Batavia (Jakarta), a pattern consistent with historical evidence that Dutch colonial development concentrated along Jalan Raya Pos and preferentially occupied the northern part of the Bandung Basin. The current urban pattern therefore retains this northward bias within the basin. The $2.4\%$ energy advantage of the southward configuration reflects the measurable difference in altitudinal gradient between the northern and southern basin boundaries. However, this lower-energy  configuration approaches the volcanic highlands that delimit the Bandung Basin. While energetically favourable, this configuration would be exposed to a combination of landslide susceptibility, volcanic hazards and geotechnical instability associated with the basin's volcanic setting, making it a physically less favorable location for long-term urban development \cite{bandung1, bandung2,bandung3}.

\paragraph{Buenos Aires ($\Delta = -0.74\%$, Supplementary Fig.~\ref{fig:supp:buenosaires_bruxelles} top).} Buenos Aires lies on the exceptionally flat Pampean plain, where topographic variation is minimal. The R\'{\i}o de la Plata estuary borders the city to the east, extending northward into the Paran\'a delta system, which renders eastward and most northward translations infeasible. This deltaic-estuarine setting forms part of a large floodplain and wetland system associated with the lower Paran\'a and R\'{\i}o de la Plata, rather than a homogeneous land surface available for unconstrained displacement. The accessible configuration space is therefore limited primarily to southward and westward directions. The best-displaced configuration is a southward translation of approximately $1750\, \mathrm{m}$. The marginal $-0.7\%$ preference likely reflects a slight north–south asymmetry: the northern metropolitan districts, proximal to the Paran\'a delta fringe, involve  more complex routing near river channels and low floodplain terrain, while the southern metropolitan area sits on uniform pampa. The deviation is close to the numerical precision threshold of the analysis and is well within the context of a nearly flat landscape \cite{ba1,ba2,ba3,ba4}.

\paragraph{Brussels ($\Delta = -0.49\%$, Supplementary Fig.~\ref{fig:supp:buenosaires_bruxelles} bottom).}

Brussels occupies the transition between the flat Flemish plain to the north and the gently rising Brabant plateau to the south and southeast, with local relief ranging from approximately 15 m in the northern districts to over 100 m in the southeastern uplands. The Sonian Forest (For\^et de Soignes, a Natura 2000 protected area) marks the southeastern urban boundary at higher elevation, contributing to the asymmetric energetic landscape. The best-displaced configuration is a northward translation of $2250\,\mathrm{m}$, which moves the road network toward the flatter Flemish terrain and reduces cumulative altitudinal costs on routes that currently cross the plain-to-plateau gradient. The $-0.49\%$ advantage reflects the moderate relief of the Brussels region and indicates that the observed city position is already well adapted to its topographic context. At the same time, the northward counterfactual should be interpreted as a geometric displacement rather than as a freely available urban alternative, since the northern and canal-side sectors of Brussels are strongly structured by railway, logistics and productive infrastructures \cite{bu1,bu2}.

\section*{Supplementary Note 5: Geographic constraints and infeasible configurations}
\label{sec3}

\begin{figure*}[p]
\centering
\includegraphics[width=\textwidth]{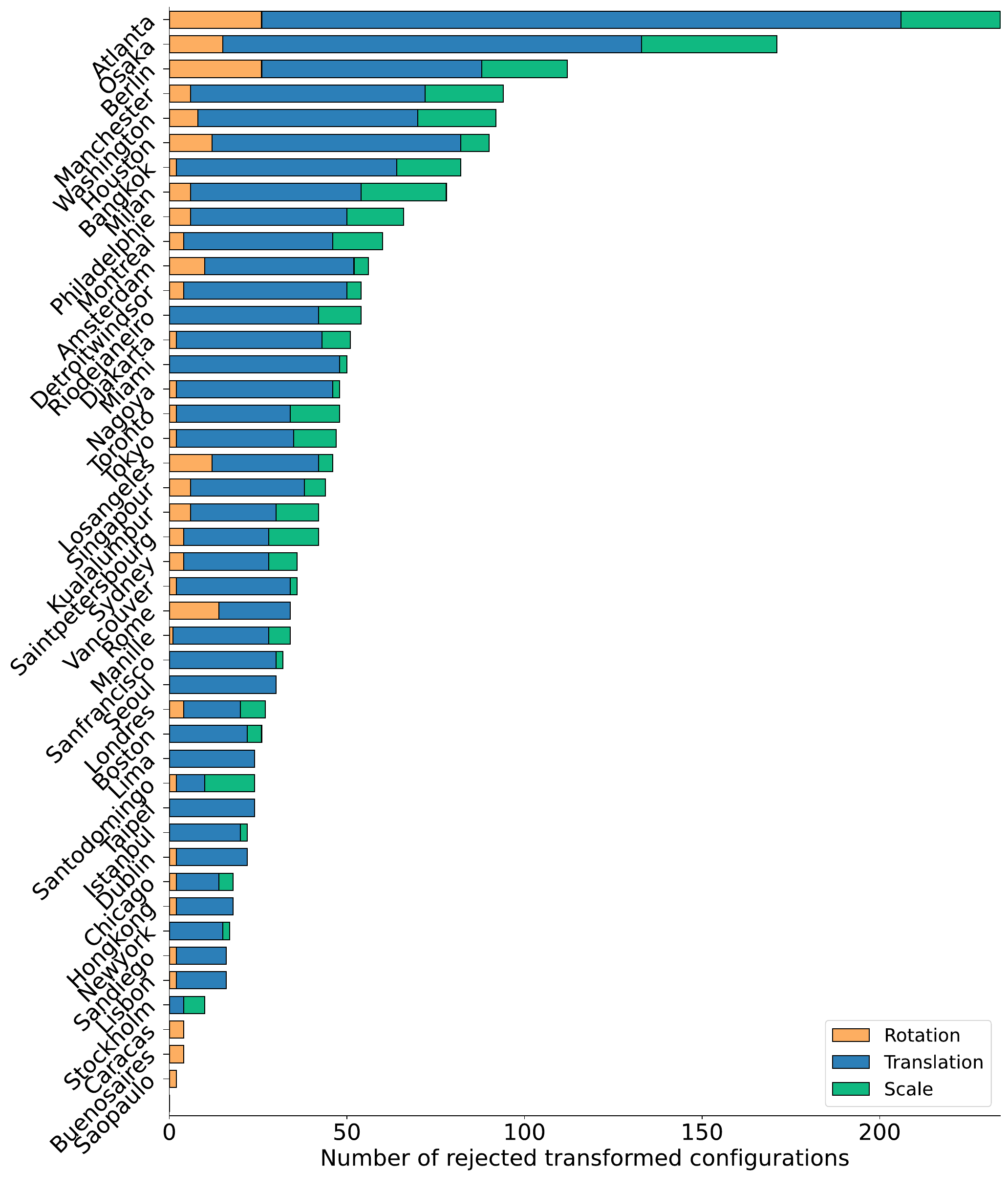}
\caption{\label{fig:supp:landcounts}\textbf{Number of rejected transformed configurations per city.} A transformation is rejected when at least one node of the road network is placed outside the valid terrestrial domain or the digital elevation model support. Bars are stacked by transformation family (rotations in orange, translations in blue, rescalings in green). City labels follow the internal dataset identifiers. Cities absent from this plot experienced no rejections and had access to the full transformation set.}
\end{figure*}

A transformed embedding is rejected if any node of the road network is placed outside the valid terrestrial domain or outside the digital elevation model support. These exclusions reflect geographic constraints, such as coastlines, lakes and reservoirs, or the limits of the elevation data domain, and they restrict the feasible configuration space of each city. Supplementary Fig.~\ref{fig:supp:landcounts} shows the number of rejected transformations per city.

The distribution of rejections is strongly non-random. Urban systems near coastlines, estuaries, island chains, lake shores, delta plains, and mountain fronts accumulate more rejected configurations than inland cities. This asymmetry directly reduces the size of the counterfactual comparison set and explains why 26 cities in our dataset are classified as geographically constrained: for these cities, a substantial fraction of the tested translations, rotations, or rescalings would place part of the network over water or outside the valid terrain domain.

Chicago provides an illustrative example of a geographically constrained city. Due to the position of its Functional Urban Area relative to Lake Michigan, virtually all transformed embeddings become infeasible. Eastward translations place network nodes over water, while the remaining transformations fail to generate physically admissible alternatives within the available domain. Consequently, no lower-energy displaced configuration can be meaningfully tested, and Chicago is classified as constrained rather than as a city with a competing lower-energy alternative.

Tokyo's functional urban area exemplifies a large coastal constraint. The metropolitan area borders Tokyo Bay to the east and south, with an extensive reclaimed shoreline built up over several centuries of land-making (from the Edo period to the present day). Eastward and southward translations rapidly place network nodes over the bay, while complex coastal geometries further reduce the fraction of admissible embeddings. The Kanto plain extends to the north and west and is accessible, but the bay imposes a sharp and extensive boundary on the other sides.

\section*{Supplementary Note 6: OSM municipality boundary analysis}
\label{sec:osm}

The main analysis uses the GHS Functional Urban Area boundaries (European Commission, JRC), based on the EU--OECD definition, which captures the full functional extent of the metropolitan system including commuter zones and suburban networks. As a sensitivity test, we repeated the analysis using the administrative municipality boundary downloaded from OpenStreetMap (OSM), which typically encompasses only the urban core. Results are compared in Supplementary Fig.~\ref{fig:supp:osm}.

\begin{figure*}[ht]
\centering
\includegraphics[width=0.96\textwidth]{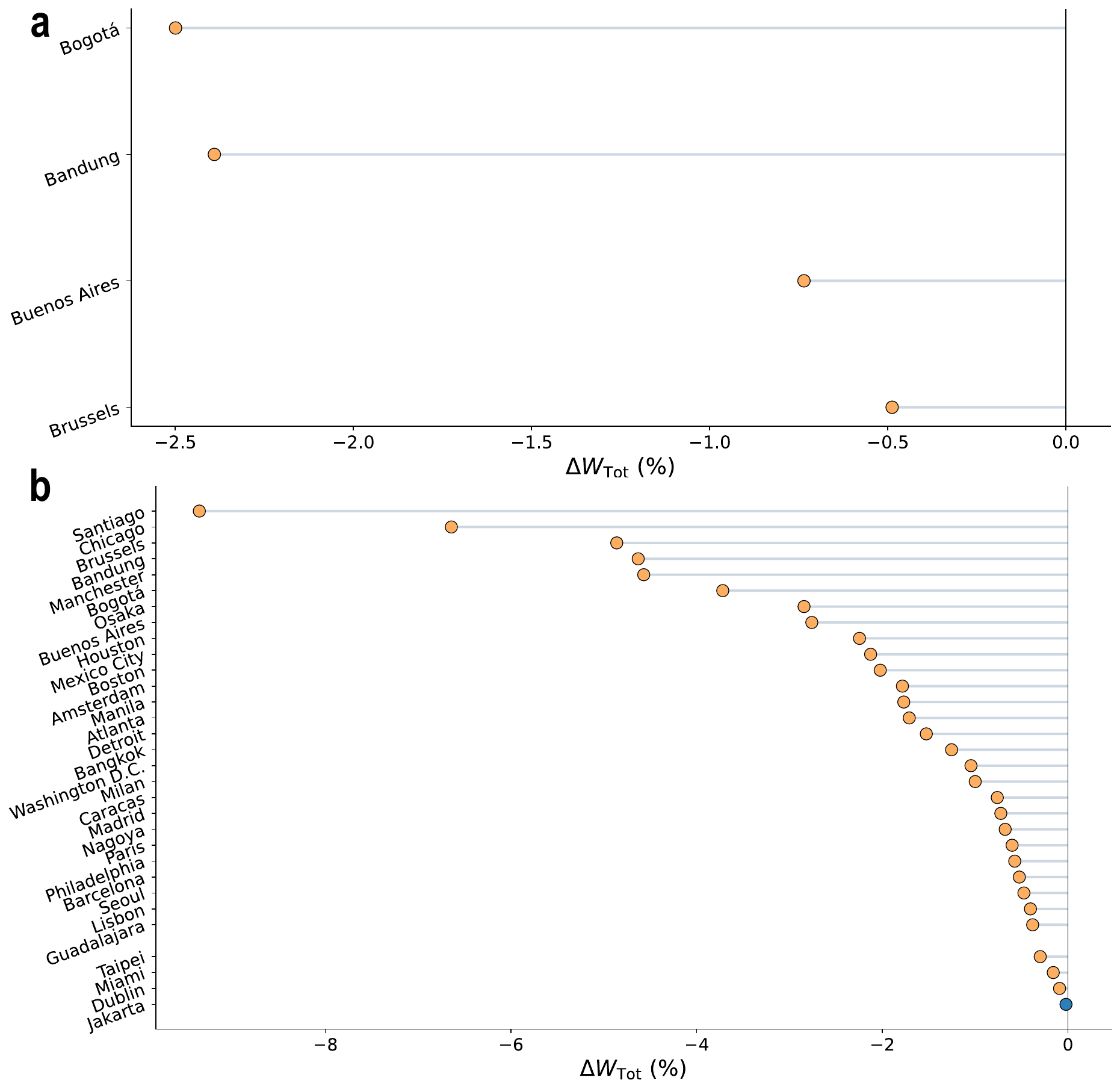}
\caption{\label{fig:supp:osm}
\textbf{Cities with lower-energy displaced configurations under different spatial domains.}
Each point represents a city for which at least one transformed embedding yields a lower total mobility energy than the empirical configuration, quantified as
$\Delta = 100\times(W_{\min}/W_{\mathrm{orig}}-1)$.
Panel (\textbf{a}) shows results obtained using Functional Urban Area (FUA) boundaries, while panel (\textbf{b}) uses OpenStreetMap municipality boundaries.
The substantially larger number of negative-energy configurations obtained under municipality boundaries highlights the strong dependence of the analysis on the spatial domain used to define the urban system.}
\end{figure*}

\subsection*{Results under OSM municipality boundaries}

Of the 56 cities in the main analysis, 55 have a usable OpenStreetMap municipality boundary; Tokyo's metropolitan area has no single administrative boundary polygon in OSM and was therefore excluded from this comparison. Under the OSM municipality boundary, 13 of these 55 analyzed cities show the original embedding as the energetic minimum, 32 cities exhibit a displaced alternative with lower $W_{\mathrm{Tot}}$, and the remaining 10 cities are geographically constrained under the municipality boundary, admitting no valid alternative transformations to compare against. The deviations are frequently larger under the municipal boundary: for example, Santiago shows $\Delta = -9.4\%$, Chicago $\Delta = -6.6\%$, and Manchester $\Delta = -4.6\%$, compared to $0\%$, constrained, and $0\%$ respectively under the FUA.

\subsection*{Why the FUA is the appropriate spatial domain}

This contrast reveals that the result depends critically on the spatial scale at which mobility is modeled. Several factors explain why the FUA provides the more appropriate spatial domain for the present mobility-based analysis:

\paragraph{Representativeness of the mobility domain.} The FUA boundary is designed to capture the full extent of the origin-destination flows that constitute daily urban mobility, including commuting patterns and suburban trips. The municipality boundary, by contrast, encloses only the urban core. Truncating the network at the municipality boundary discards a large fraction of actual trips and artificially concentrates the network near topographic features---hills, rivers, coastal zones---that form the edges of historic city centers.

\paragraph{Boundary effects on the energetic landscape.} When the network is small (municipality scale), a geometric displacement of a few kilometers can move a large fraction of nodes close to a topographic feature (e.g., a river valley, a coastal plain), producing apparent energetic gains that are simply artifacts of the boundary choice. At the FUA scale, the network is large enough that the same displacement has a proportionally smaller effect on the average exposure to terrain.

\paragraph{Functional vs.\ administrative delineation.} The municipality boundary is an administrative construct that reflects historical city limits, not the actual extent of daily mobility. Cities such as Santiago, Madrid, or Paris extend far beyond their administrative boundaries in terms of daily commuting patterns. Using the municipality boundary systematically underestimates the spatial scale over which the mobility--terrain interaction operates.

\paragraph{Sensitivity of the FUA result.} The consistency of the FUA result across cities with different geographies, histories and administrative structures suggests that the low-energy pattern is more clearly expressed at the functional
urban scale. The OSM municipality result, by contrast, varies substantially across cities in ways that correlate with the degree to which the municipality underrepresents the true functional urban area, rather than with physical geography.

In summary, the OSM municipality analysis indicates that the low-energy configuration pattern is scale-dependent: it emerges at the functional urban scale and is not visible at the administrative municipality scale. This scale dependency is itself physically informative: it suggests that the alignment between urban form and terrain is a property of the metropolitan system as a whole, not of the administrative core in isolation.

\section*{Supplementary Note 7: DEM product validation}

To verify that the SRTM-based digital elevation model used throughout the analysis does not introduce systematic bias, we compared node-level elevation values from the SRTM product against corresponding values extracted from the Copernicus DEM (GLO-30) for two cities spanning contrasting relief regimes: Barcelona (moderate, coastal topography) and Santiago (high-relief, Andean setting).

\begin{figure*}[ht]
\centering
\includegraphics[width=0.46\textwidth]{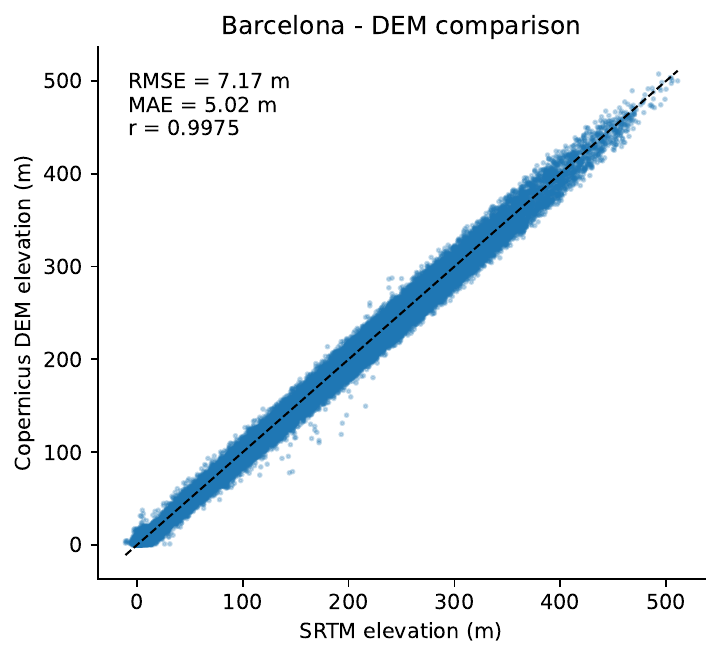}\hfill
\includegraphics[width=0.46\textwidth]{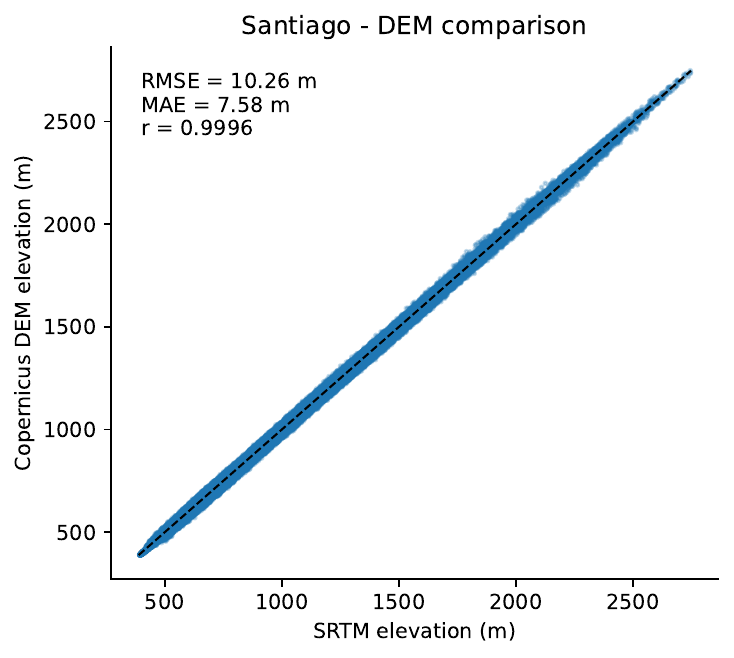}
\caption{\label{fig:supp:demcorr}\textbf{Comparison of SRTM-based and Copernicus-based elevation values at road-network nodes.} Each point represents one node. The dashed line is the identity ($y=x$). Left: Barcelona (relief range 0--600\,m). Right: Santiago (relief range 500--2\,800\,m). In both cities the correlation is near-perfect, indicating that the choice of elevation product does not materially affect node elevations.}
\end{figure*}

In both cities the two datasets are in close agreement across the full elevation range, with the scatter tightly concentrated around the identity line (Supplementary Fig.~\ref{fig:supp:demcorr}). This confirms that the choice of DEM product does not materially affect node elevations.

\section*{Supplementary Note 8: Contraction (shrink) transformations}
\label{sec:shrink}

The main text (Methods, City replicas) restricts the anisotropic scaling ensemble to expansions of the built area ($1.02$--$1.20$) and does not test contractions, on the grounds that shrinking a network compresses its nodes and edges into a footprint smaller than the real one so that any energy reduction it produced would simply reflect the shorter paths created by the compression itself, rather than a more favorable alignment between the network and the terrain. Here we test that reasoning directly, for the two cities used elsewhere as the main systematic-transformation showcase (Madrid and Santiago), by generating the mirror-image contraction factors of the expansion set already used in the paper: $0.98,\,0.95,\,0.92,\,0.90,\,0.88,\,0.85,\,0.80$, applied independently along the north--south and east--west axes with the same pipeline used for every other transformation in this study.

\begin{figure*}[ht]
\centering
\includegraphics[width=0.96\textwidth]{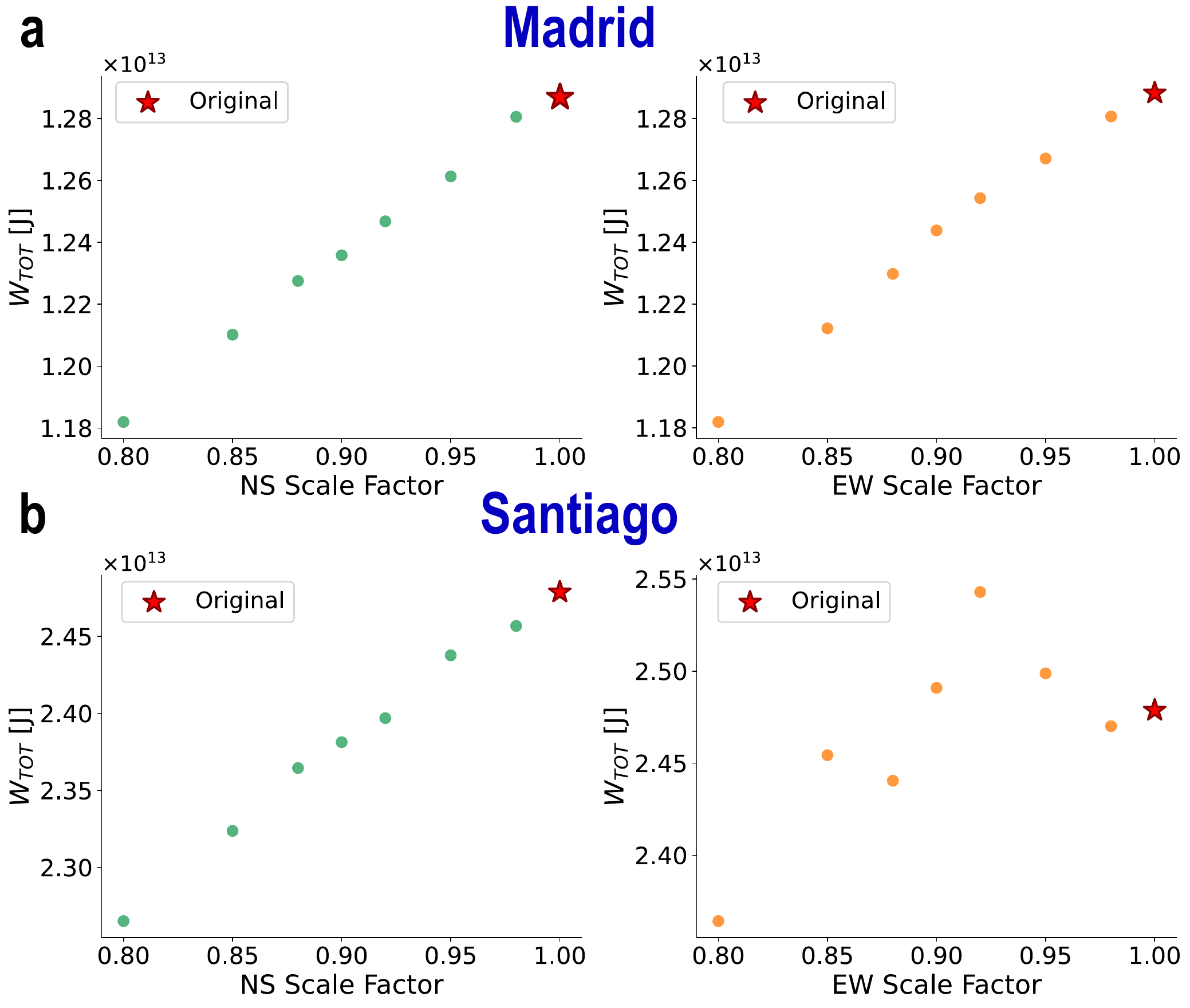}
\caption{\label{fig:supp:shrink}\textbf{(a)} Madrid and \textbf{(b)} Santiago --- energy profile under contraction (shrink) scaling. Total mobility energy $W_{\mathrm{Tot}}$ as a function of the north--south (left) and east--west (right) contraction factor, from $0.80$ to $0.98$. The red star marks the empirical (original, factor $=1.0$) configuration.}
\end{figure*}

Supplementary Figs.~\ref{fig:supp:shrink}a and \ref{fig:supp:shrink}b show the resulting energy profiles. In Madrid along both axes and in Santiago along the north--south axis, $W_{\mathrm{Tot}}$ decreases essentially monotonically as the scaling factor moves away from $1.0$ towards $0.80$. Santiago's east--west profile is non-monotonic: intermediate factors ($0.90$--$0.95$) yield energies above the original, and the decrease appears only for stronger contractions ($0.80$--$0.88$), indicating that terrain alignment can also modulate the response along this axis. Overall, the energy reduction under strong contraction follows the shrink factor, as expected from the compression artifact described above, rather than revealing a configuration that is both more compact and better aligned with the terrain. This is consistent with, though does not on its own prove, our decision to exclude contraction from the transformation ensemble reported in the main text.

\end{document}